\documentclass[published, twocolumn, 12pt]{aastex63}

\def\lsim{~\rlap{$<$}{\lower 1.0ex\hbox{$\sim$}}}
\def\gsim{~\rlap{$>$}{\lower 1.0ex\hbox{$\sim$}}}

\begin{document}

\title{\large Next Generation Very Large Array Memo No. 66:\\ Exploring Regularized Maximum Likelihood Reconstruction\\for Stellar Imaging with the ngVLA}

\author[0000-0002-9475-4254]{Kazunori Akiyama}
\affiliation{National Radio Astronomy Observatory, 520 Edgemont Road, Charlottesville, VA 22903, USA}
\affiliation{Massachusetts Institute of Technology Haystack Observatory, 99 Millstone Road, Westford, MA 01886, USA}
\affil{National Astronomical Observatory of Japan, 2-21-1 Osawa, Mitaka, Tokyo 181-8588, Japan}
\affil{Black Hole Initiative, Harvard University, 20 Garden Street, Cambridge, MA 02138, USA}

\author[0000-0002-3728-8082]{Lynn D. Matthews}
\affiliation{Massachusetts Institute of Technology Haystack Observatory, 99 Millstone Road, Westford, MA 01886, USA}

\begin{abstract}
The proposed next-generation Very Large Array (ngVLA) will enable the imaging of astronomical sources in unprecedented detail by providing an order of magnitude improvement in sensitivity and angular resolution compared with radio interferometers currently operating at 1.2--116~GHz.
However, the current ngVLA array design results in a highly non-Gaussian dirty beam that may make it difficult to achieve high-fidelity images with both maximum sensitivity and maximum angular resolution using traditional \texttt{CLEAN} deconvolution methods.
This challenge may be overcome with regularized maximum-likelihood (RML) methods, a new class of imaging techniques developed for the Event Horizon Telescope. RML methods take a forward-modeling approach, directly solving for the images without using either the dirty beam or the dirty map. 
Consequently, this method has the potential to improve the fidelity and effective angular resolution of images produced by the ngVLA. As an illustrative case, we present ngVLA imaging simulations of stellar radio photospheres performed with both multi-scale (\texttt{MS-}) \texttt{CLEAN} and RML methods implemented in the \texttt{CASA} and \texttt{SMILI} packages, respectively.
We find that both \texttt{MS-CLEAN} and RML methods can provide high-fidelity images recovering most of the representative structures for different types of stellar photosphere models. 
However, RML methods show better performance than \texttt{MS-CLEAN} for various stellar models in terms of goodness-of-fit to the data, residual errors of the images, and in recovering representative features in the ground truth images.
Our simulations support the feasibility of transformative stellar imaging science with the ngVLA, and simultaneously demonstrate that RML methods are an attractive choice for ngVLA imaging.
\vspace{4eM} 
\end{abstract}

\section{Introduction\protect\label{intro}}
The next-generation Very Large Array (ngVLA) has been conceived to enable transformative science across a broad range of astrophysical topics by providing an
order of magnitude improvement in sensitivity and angular
resolution compared with radio interferometers currently operating in the 1.2--116~GHz frequency range \citep{Selina2018}. 
Details of the ngVLA design are being informed by the requirements of designated
``key science goals'' \citep{Murphy2018}, and
addressing the diverse needs of
these science programs will require both high angular
resolution and excellent 
surface brightness sensitivity. Because the ngVLA will be non-configurable, 
this will necessitate an array with  baselines 
that  sample a
wide range of angular scales.
The currently proposed ngVLA design\footnote{See
  {\url{https://ngvla.nrao.edu/page/tools}}.} calls
for a heterogeneous array of 244 antennas of 18~m diameter and 
19  dishes with 6~m diameters \citep{Selina2018}. The smaller dishes will be confined to a
``Short Baseline Array'' with baselines of 11--56~m, to be used for
total power measurements and mapping extended and/or low surface
brightness emission.  The ``Main
Array'' will comprise 214 of the 18~m antennas on baselines
ranging from tens of meters to $\sim$1000~km. Finally, 30 
of the 18~m antennas will be distributed in a ``Long
Baseline Array'' spread
across the North American continent, with baselines up to 8860~km to
be used for Very Long Baseline Interferometry (VLBI).  

\begin{figure}[t]
\centering
\includegraphics[width=0.8\columnwidth]{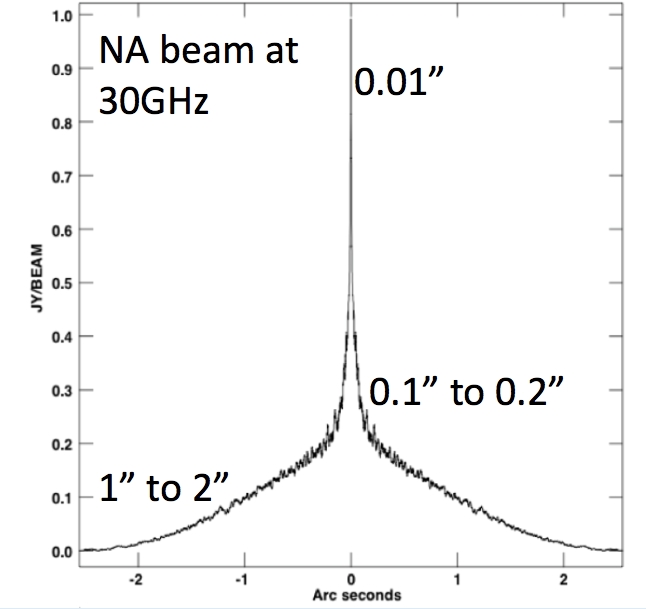}
\caption{Naturally weighted point spread function (dirty beam) for the ngVLA Main Array at 30~GHz. Reproduced from \citet{Carilli+2018}.
\label{fig:carilli_beam}}
\end{figure}

In the current ngVLA design, 
the Main Array is
``tri-scaled''  \citep[e.g.,][]{Carilli2017, Carilli2018}, 
comprising:
(i) a densely sampled, 1~km-diameter core of 94 antennas; (ii) a
VLA-scale array of 74 antennas with baselines out to $\sim$30~km; (iii) 
extended baselines (46 stations) out to $\sim$1000~km.
While in principle the Main Array is well-suited to meeting the
requirements of the ngVLA
for angular resolution, point source sensitivity, and surface
brightness sensitivity
(while respecting geographical
considerations), the antenna distribution of the Main Array results
in a highly non-Gaussian dirty beam. With natural weighting, its shape comprises a narrow core atop a
two-tiered ``skirt'' \citep[Figure~\ref{fig:carilli_beam};  see also][]{Carilli2017, Carilli2018, Carilli+2018}.
This poses a challenge for imaging ngVLA data with traditional \texttt{CLEAN} deconvolution methods, 
in which a model of the ideal ``\texttt{CLEAN} beam'' is determined by
fitting a Gaussian to the dirty beam point spread function \citep[e.g.,][]{Hogbom74}.  A consequence  is that
it is difficult to achieve maximum angular resolution in an ngVLA \texttt{CLEAN} image without sacrificing
sensitivity \citep{Carilli2017, Carilli2018, Rosero2019}. 
This problem cannot be overcome through the application 
of robust weighting \citep{Briggs99} during the deconvolution \citep{Carilli2017},
and it currently poses a potential inherent
limitation to the array performance. 

Here we present the results of a pilot study aimed at exploring the effectiveness of an alternative imaging methodology known as {\em regularized maximum likelihood} methods \citep[RML methods; see][for an overview]{EHTC2019d} for ngVLA imaging applications. As illustrative test cases, we focus on several examples of relevance to the problem of resolved imaging of stellar photospheres at radio wavelengths. For these test cases we quantitatively and qualitatively evaluate simulated ngVLA images of model stellar sources obtained using RML methods and compare the results to traditional \texttt{CLEAN} deconvolution.

In the sections that follow, we first provide an introduction to RML imaging methods and briefly review their applications to astronomical imaging to date (Section~\ref{RMLintro}).  We then discuss as a sample science application the imaging of stellar radio photospheres (Section~\ref{stars}) and undertake the computation of simulated observations of radio photospheres with the ngVLA Main Array (Section~\ref{modelsandsims}). In Section~\ref{imaging} and \ref{results}, we present the results of our imaging tests based on both RML and traditional \texttt{CLEAN} methods and present a comparative analysis of the results. A summary and future prospects are presented in Section~\ref{discussion}.

\section{RML Methods\protect\label{RMLintro}}
A recent acceleration in the development of RML imaging methods has been motivated 
by the needs of the millimeter (mm) VLBI community, 
including the Event Horizon Telescope \citep[EHT;][]{EHTC2019b} and their goal of imaging the shadows of supermassive black holes.
This goal requires improved imaging techniques that can overcome various technical hurdles (see overview by \citealt{Fish2016}).
In the case of the EHT, the primary
challenges are: 
(1) reconstructing high-fidelity images for 
sources that have complicated structure on scales comparable to the
angular resolution;
(2) reconstructing images from data with many
residual calibration errors; and (3) imaging intrinsically
time-variable emission structures \citep[see][]{EHTC2019b}.
 
This new generation of imaging techniques directly solves the
interferometric imaging equations, with single (or combinations of) regularization functions based on different prior information that enables selection of a conservative image from an infinite number of possible images providing reasonable fits to the data \citep{EHTC2019d}. Popular classes of RML techniques are {\it sparse modeling}---enforcing sparsity in some basis of the image \citep[e.g.][]{Honma2014, Ikeda2016, Akiyama2017a, Akiyama2017b, Kuramochi2018}, and {\it maximum entropy} methods \citep{Chael2016}---maximizing the information entropy of the image.

In the framework of RML methods, many observing effects attributed to observational equations (for instance, both thermal and systematic errors in the data), can be flexibly incorporated into likelihood terms of the imaging equations. Furthermore, RML methods allow direct use of robust closure quantities, free from the
calibration errors of each interferometer station \citep[e.g.,][]{Lu2014, Bouman2016, Chael2016, Chael2018, Akiyama2017a}. In addition, these methods can be used to dynamically solve for images of a time-varying target \citep{Johnson2017,Bouman2018}. 

One advantage of RML imaging methods is that because they fit the
visibility data directly, it is possible to avoid image errors
inherent to deconvolution of the dirty beam, as is done in \texttt{CLEAN} \citep{Honma2014}. For EHT imaging applications, these new methods
have been shown to consistently outperform traditional \texttt{CLEAN} and provide high-fidelity imaging even on spatial scales a factor
of $\sim$3--4 smaller than the nominal diffraction limits (i.e., they allow for modest super-resolution of the data) --- without the artifacts inherent to super-resolved \texttt{CLEAN} images. 

Overall RML methods provide a more flexible framework of interferometric imaging than conventional \texttt{CLEAN} techniques, with a higher performance particularly on high-angular-resolution imaging. This combination of properties makes these new imaging methods potentially well-suited to
the imaging needs of the ngVLA.

\begin{figure*}[ht]
\centering
\gridline{
    \fig{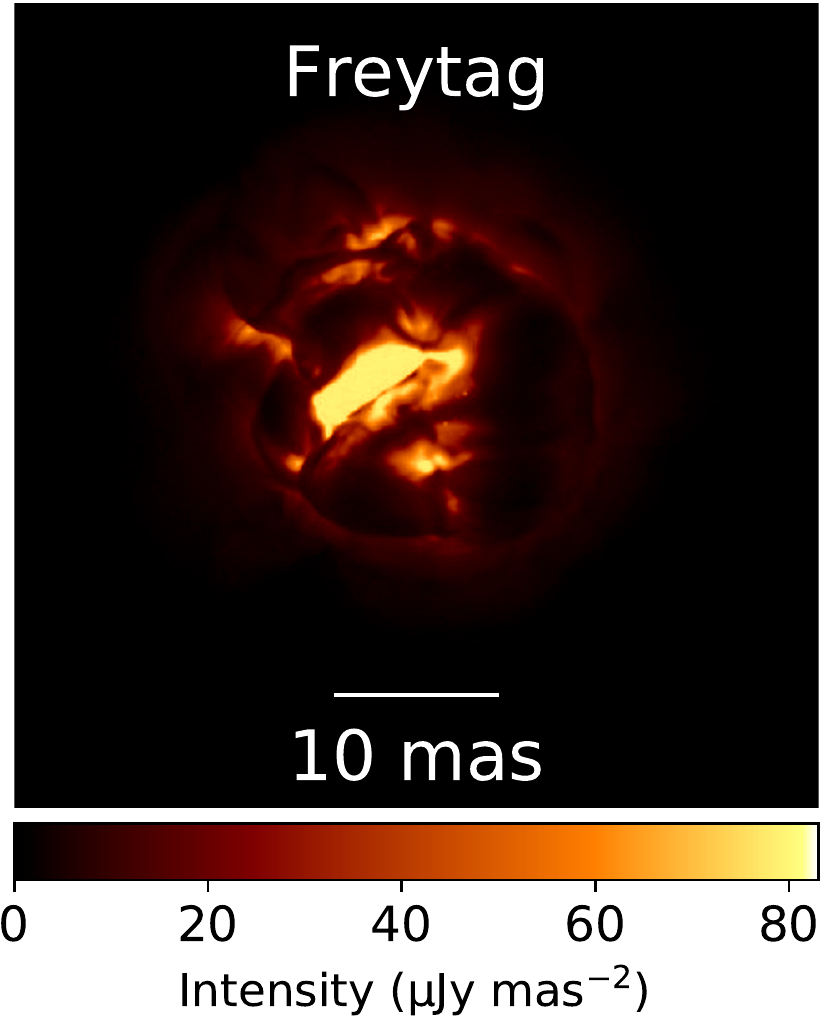}{0.23\textwidth}{(a) \texttt{Freytag} model}
    \fig{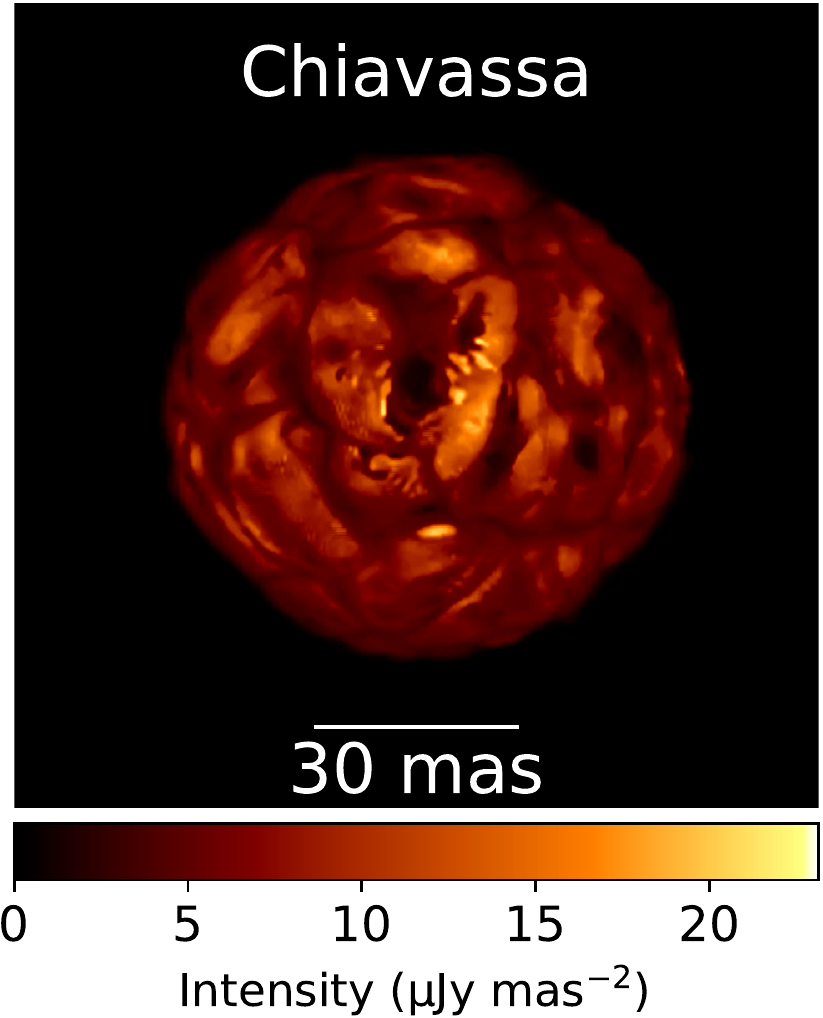}{0.23\textwidth}{(b) \texttt{Chiavassa} model}
    \fig{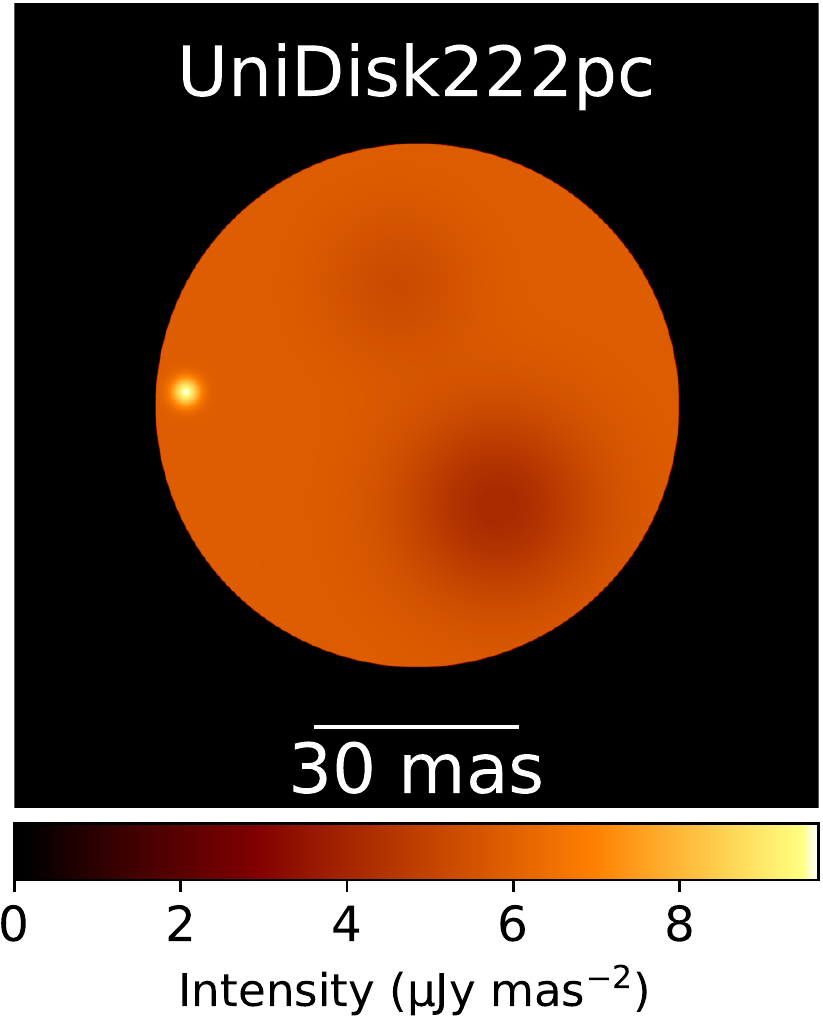}{0.23\textwidth}{(c) \texttt{UniDisk222pc} model}
    \fig{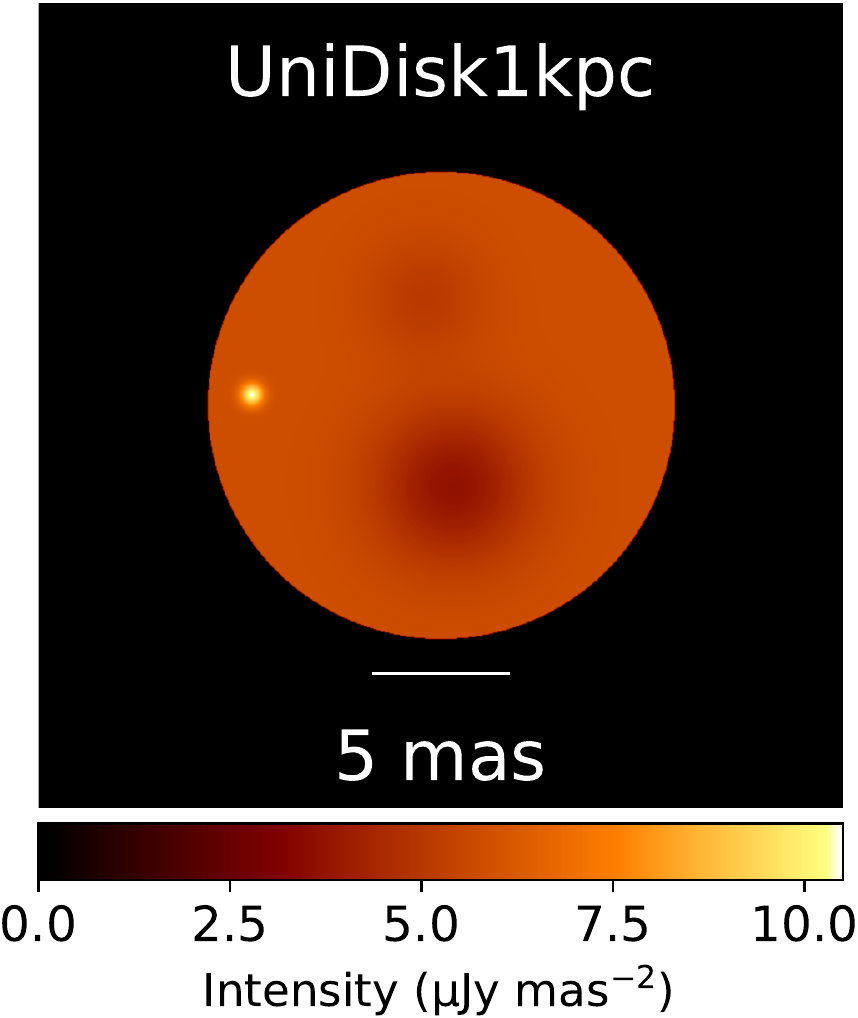}{0.24\textwidth}{(d) \texttt{UniDisk1kpc} model}
}
\caption{
The four stellar models adopted in the current study. See Section \ref{modelsandsims} for details.
}
\label{fig:model_images}
\end{figure*}

\section{Imaging Stellar Radio Photospheres: a Test Case for RML Methods\protect\label{stars}}
One of the many groundbreaking scientific applications of the ngVLA will be its ability to obtain resolved images of the surfaces of nearby stars spanning a range of spectral types and evolutionary phases, from dwarfs to supergiants \citep{Carilli2018ngvla, MattClaussen2018, Harper2018}. Such observations are expected to revolutionize our ability to use radio observations as a tool in stellar astrophysics.

For asymptotic giant branch (AGB) and red supergiants (RSG) stars---whose enormous radio photospheres can span several au and subtend up to a few tens of mas---it is currently possibly to marginally resolve some of the closest ($d\lsim$150~pc) examples using the Karl G. Janksy Very Large Array (VLA) and the Atacama Large Millimeter/submillimeter Array (ALMA) in their longest baseline configurations \citep[e.g.,][]{Lim1998, RM97, RM07, Matthews2015, Matt+2018, Menten2012, OGorman2015, Vlemmings2019}. However, the ngVLA will supply an enormous leap forward in our ability to measure the detailed properties of radio photospheres (e.g., the presence of atmospheric temperature gradients, spots, and surface features, as well as temporal changes) for such stars out to $\sim$1~kpc \citep{MattClaussen2018}. Such measurements will supply unique insights into the atmospheric physics, including the temperature structure of the atmosphere, and constraints on the mechanisms (e.g., shocks, pulsation, convection) that help to drive the observed high rates of mass loss from these stars. 
Such observations will also enable for the first time, detailed comparisons with predictions of state-of-the-art dynamic 3D atmospheric models of AGB stars and RSGs that are just now becoming possible with modern supercomputers. 

Modeling the dynamic atmospheres of AGB stars and RSGs is  extraordinarily
challenging owing to their complex
physics and non-LTE conditions. However, the latest generations of 3D models now incorporate the effects  pulsation, convection, shocks, and dust condensation and provide detailed 
predictions with high time and spatial resolution \citep[e.g.,][]{Chiavassa2009, Freytag2017, Liljegren2018}.   As a next step, empirically testing the exquisitely
detailed predictions of these new models will demand
new ultra high-resolution measurements of diverse samples of
stars using instruments like the ngVLA. 

Recently \cite{Matt+2018} presented the first tests of RML methods for imaging the radio photospheres to a sample of nearby ($d\lsim$150~pc) AGB stars observed with the VLA at 46~GHz. 
Because even the nearest AGB stars are only marginally resolved by the
current VLA, \texttt{CLEAN} images can reveal evidence of non-circular shapes, but provide little or no information
on the possible presence of predicted photospheric features such as giant convective cells \citep{S75} or other temperature non-uniformities.
In contrast, the modest degree of super-resolution enabled by use of RML imaging methods supplied evidence for the first time of non-axisymmetric shapes
and/or non-uniform surface brightnesses in all five of the sample stars.

In the current study we investigate how the attributes of RML imaging can be exploited to address the new set of imaging challenges inherent to the current ngVLA design (see Section~\ref{intro}), using stellar imaging as an illustrative test case. It is anticipated that applications of RML methods to imaging other classes of sources with the ngVLA will be explored in future studies.

\begin{figure*}[t]
\centering
\gridline{
    \fig{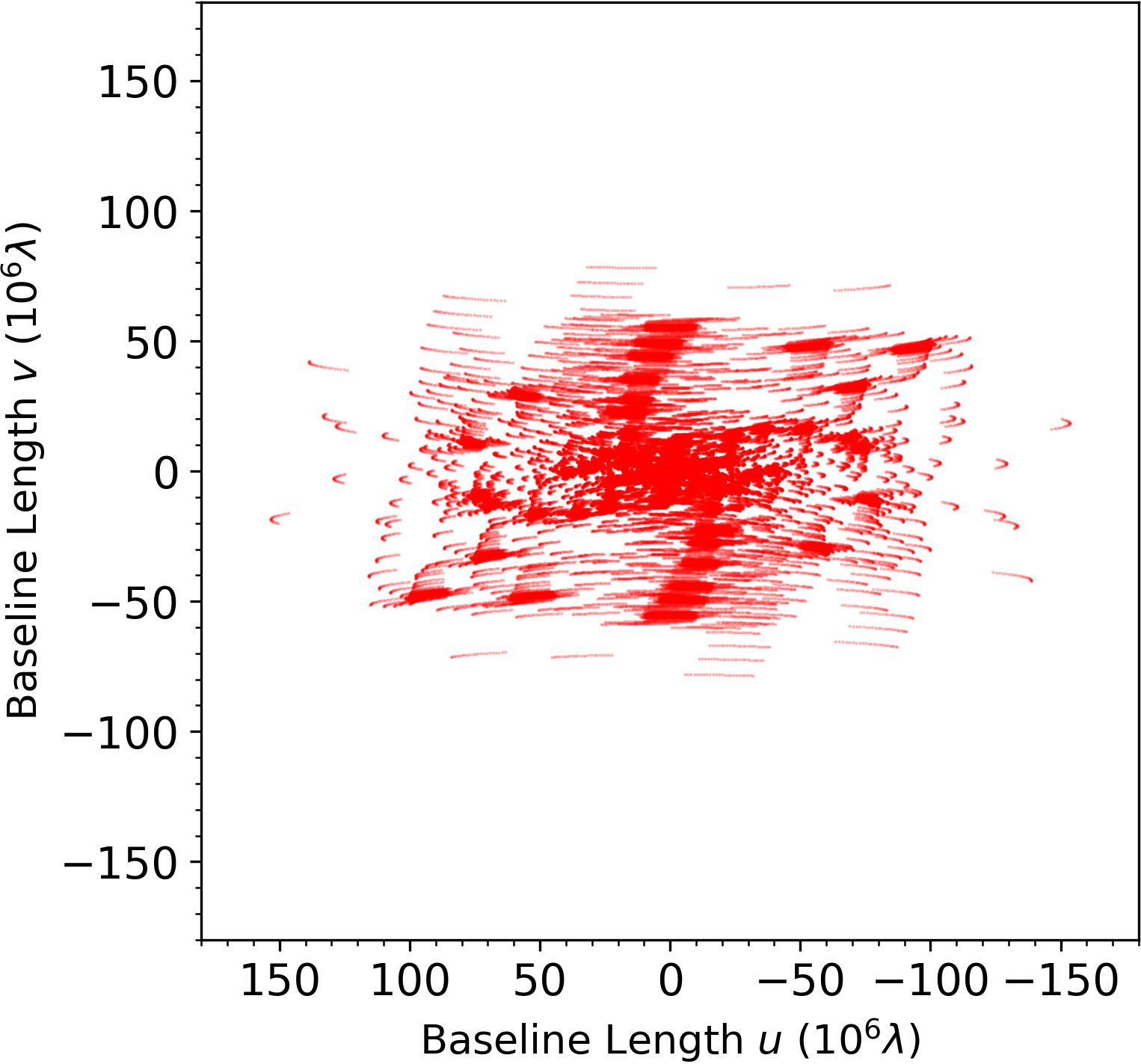}{0.33\textwidth}{(a) \texttt{Freytag} model}
    \fig{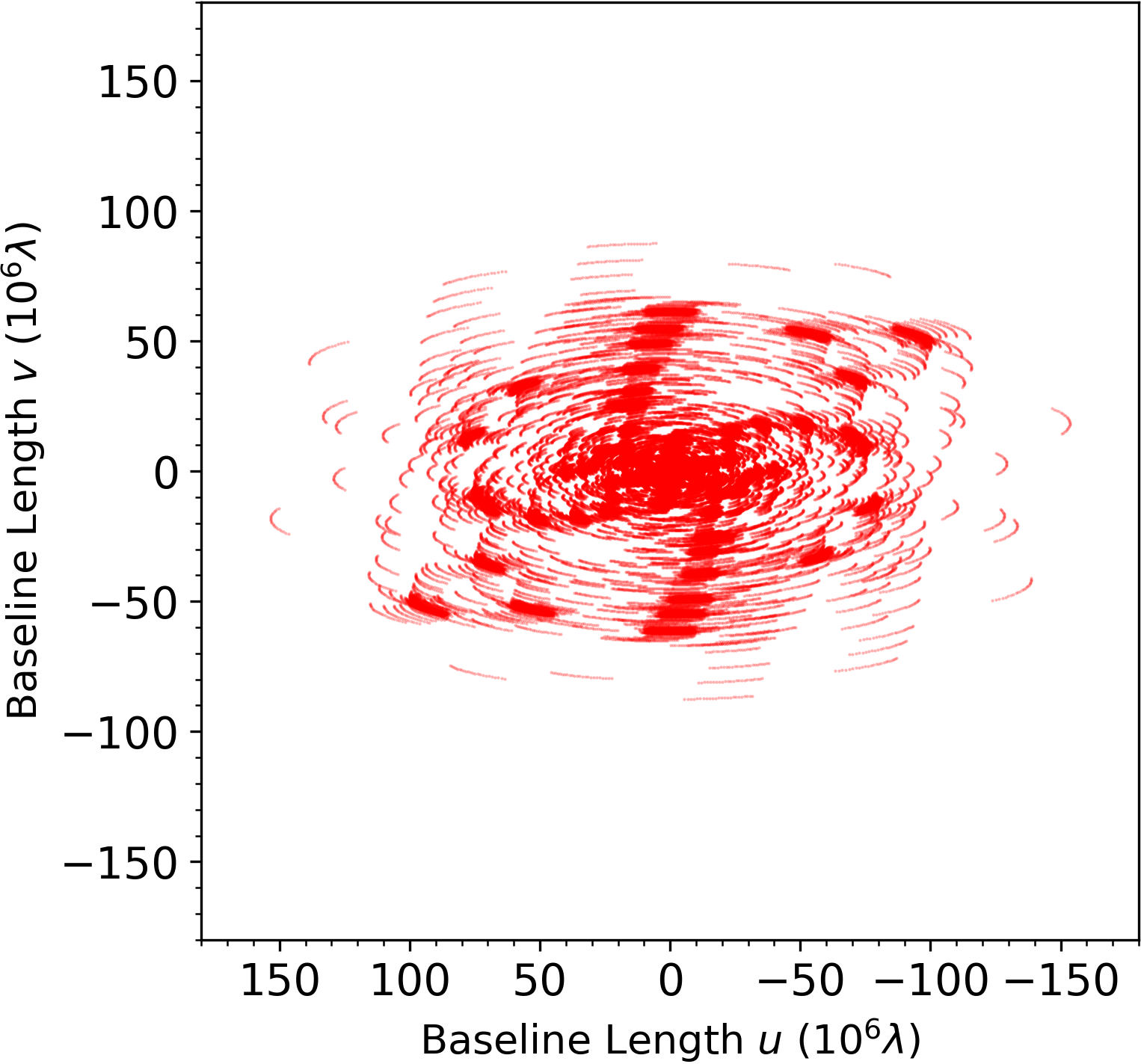}{0.33\textwidth}{(b) \texttt{Chiavassa} model}
    \fig{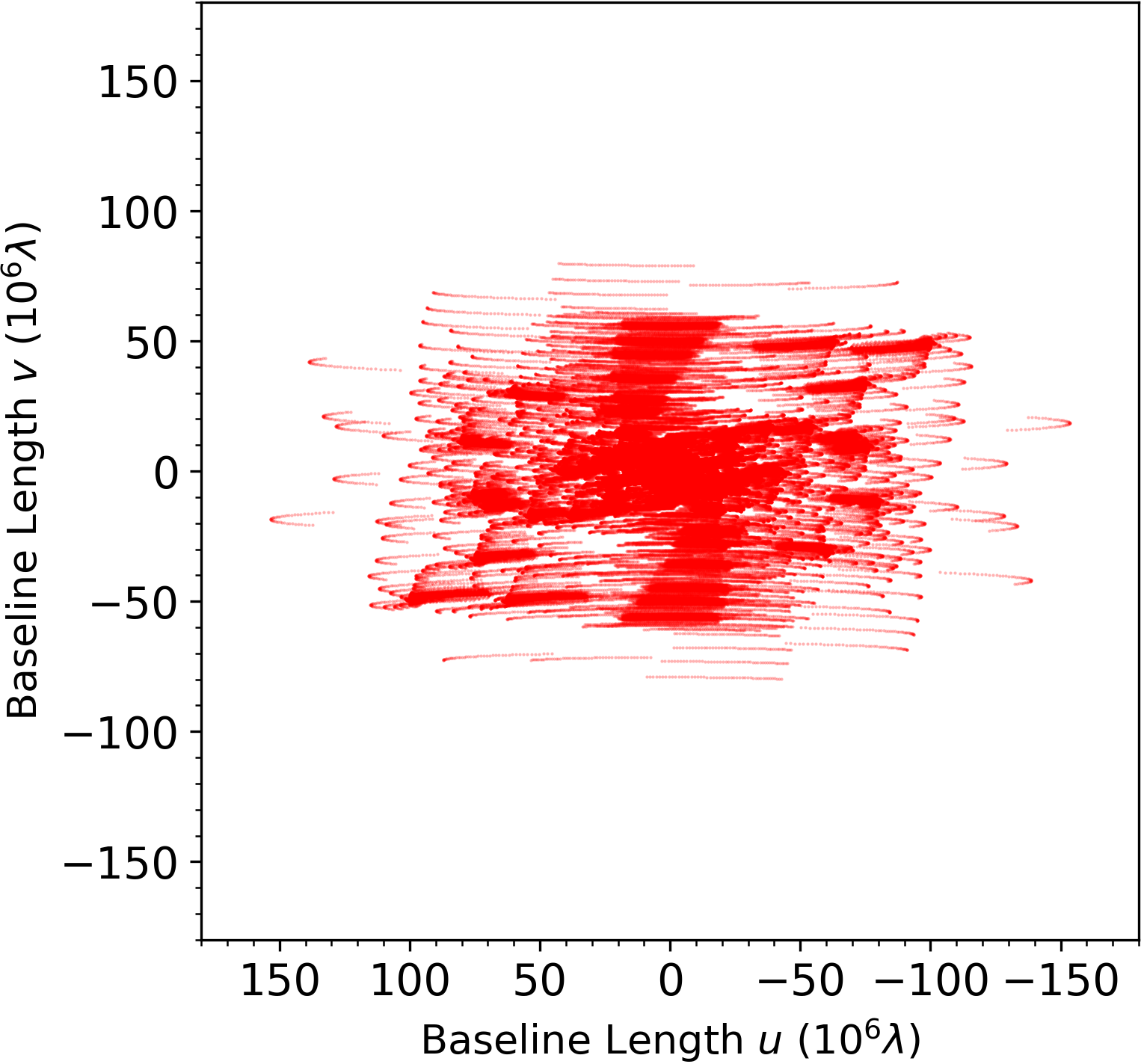}{0.33\textwidth}{(c) \texttt{UniDisk222pc} and \texttt{UniDisk1kpc} models}
}
\caption{
$uv$-coverage of simulated observations. See Section \ref{modelsandsims} for details. 
}
\label{fig:uvcoverage}
\end{figure*}

\section{Models and Simulated Observations\protect\label{modelsandsims}}
The imaging tests for the current study were performed using a series of three different simulated data sets (Figure~\ref{fig:model_images}; see also below). In each case, a ``ground truth" model photosphere was first devised and converted into one or more FITS images. The headers of the FITS images were edited to insert appropriate source coordinates, pixel scales, intensity scaling, and other crucial information as necessary. A pixel size of 0.04~mas was used in all of the ground truth images (a factor of $\sim$25 times smaller than the angular resolution of the ngVLA Main Array at 46~GHz). To avoid edge effects, zero padding was used to create a field-of-view for each ground truth frame of $\sim$0.33 arcsec per side. 

Simulated ngVLA observations of each model were performed using the \texttt{simobserve} task in \texttt{CASA} to produce model visibility data in uvfits format. In all cases the array configuration was taken to be the ngVLA Main Array (see Section~\ref{intro}), as defined in the configuration file ngvla-main-revC.cfg (see Footnote~1). Appropriate thermal noise was added to visibility data generated by \texttt{simobserve} using the prescription outlined in \cite{Carilli+2017a}. The mock observations ranged from 2--4 hours in duration and were assumed be centered on the time of the source transit. The resulting $uv$ coverage for each observation is shown in Figure~\ref{fig:uvcoverage}. 

All of our simulations assumed dual polarizations (resulting in Stokes~I images) and a center observing frequency of 46.1~GHz ($\lambda\approx$7~mm). While the ngVLA is expected to operate at wavelengths as short as 3~mm, our adopted frequency allows direct comparisons with both real and simulated observations from the current VLA. For noise calculation purposes, we assume a total bandwidth of 10~GHz per Stokes (half the nominal value expected for the ngVLA; see \cite{Selina2018}). 

Imaging of the model visibility data sets is discussed in Section \ref{imaging}. We note that for simplicity, our current simulations are limited to a single frequency channel and thus do not attempt to evaluate the effects of multi-frequency synthesis on the image properties.

\subsection{Uniform Disk Models\protect\label{unidisk}}
To first order, a uniform disk (either circular or elliptical) is generally found to provide a satisfactory representation of the brightness distribution of the radio photospheres of nearby AGB and RSG stars as observed with current VLA and ALMA resolutions of $\sim$20--40~mas \citep[e.g.,][]{Lim1998, RM97, RM07, Menten2012, Matthews2015, Matt+2018, Vlemmings2017}. As a simple first test case, we have therefore created a model radio photosphere comprising a uniform circular disk with three ``spots" of different sizes superposed (one brighter than the underlying photosphere and two that are cooler). Such a model is useful for: (i) testing the ability of RML methods and \texttt{MS-CLEAN} to recover smooth, spatially extended emission,; (ii) testing how well the two methods can image sources with sharp boundaries; (iii) evaluating how accurately the properties of radio photospheres (including the presence of surface features) can be discerned on stars as a function of increasing distance.

For our ground truth model (\texttt{UniDisk222pc}) we adopted parameters similar to those of the RSG star Betelgeuse as measured with the VLA at 7~mm by \cite{Lim1998}. We thus assume a uniform (circular) disk diameter of 80~mas and a flux density at 46.1~GHz of 28.0~mJy.  We model the three spots as superposed circular Gaussian components with FWHM sizes and flux densities, respectively, of 24~mas ($-0.112$~mJy), 18~mas ($-0.224$~mJy), and 3.6~mas (0.056~mJy). We assume a distance for the base model star of 222~pc and created an additional version appropriately scaled to a distances of 1~kpc (\texttt{UniDisk1kpc}). These models were created using simulator tasks within the \texttt{CASA} 
toolkit.\footnote{See,\,e.g.,\,\url{https://casaguides.nrao.edu/index.php?title=Simulation_Guide_Component_Lists_(CASA_4.1)}.} The sources were assumed to have J2000 sky positions of RA=02$^{\rm h}$ 00$^{\rm m}$, DEC=$-02^{\circ}$ 00$'$; this position was intentionally chosen to result in a slightly elliptical dirty beam. 

\subsection{Making Movies of Stars: Simulations Based on Dynamic 3D Atmospheric Models}
One of the outputs of the 3D hydrodynamic simulations of AGB and RSG star atmospheres described above are ``movies" of parameters such as temperature, density, and emergent intensity as a function of time that illustrate the changes in the photospheric shape, size, brightness distribution, etc. that are predicted to occur on timescales ranging from days to several years. As shown below, the ngVLA will have the power to produce analogous movies based on real stars.

To our knowledge, none of the hydrodynamic models published to date have attempted to predict the detailed appearance of an AGB or RSG star specifically at mm wavelengths. Nonetheless, there is growing evidence based on recent VLA and ALMA imaging that radio photospheres are time-variable and non-uniform in surface brightness and that they may echo (at least to some degree), the complex and time-varying appearance of the star expected at infrared and shorter wavelengths \citep[e.g.,][]{OGorman2015, Matthews2015,  Matt+2018, Vlemmings2019}, with the caveat that the empirical information available is extremely limited owing to a combination of the limited spatial resolution and temporal coverage of the current observations. Thus the correlation between the appearance of the radio photosphere the photospheric features of the star at  shorter wavelengths is presently poorly constrained.

Despite these uncertainties, we aim here to explore the scope what will become possible with the ngVLA and to provide challenging test cases for our present imaging experiments. We have therefore adopted predictions from the existing 3D intensity models as proxies for the time-varying morphology of radio photospheres. Below we explore two examples (a nearby AGB and a nearby RSG star) that showcase the ngVLA's expected ability to make  extraordinarily detailed movies of evolving radio photospheres. 

In formulating our ground truth models, we use the results of existing 46~GHz measurements \citep[e.g.,][]{Lim1998, RM07} to set the size and mean brightness temperature of the two model stars. However, we caution that finer details of these models, such as the minimum and maximum brightness temperature, the size scales of the observed surface features, and the magnitude of the temporal variations, should be regarded as merely illustrative.

\subsubsection{Model of an Evolving AGB Star\protect\label{Freytag}}
To simulate the appearance of the time-varying radio photosphere of a 1~$M_{\odot}$ AGB star we have adopted model st28gm06n25 from \cite{Freytag2017}. This model has a bolometric luminosity $L=6890~L_{\odot}$, a mean  effective temperature $T_{\rm eff}$=2727~K, and a pulsation period $P$=1.388~yr. Freytag et al.'s model calculation was performed within a box spanning 1970$R_{\odot}$ per side ($\sim$9.1~AU). We assume that the star is at a distance of $\sim$150~pc and that at 46.1~GHz it subtends a mean angular diameter of $\sim$50~mas and has an integrated flux density of 10~mJy.

To create a ground truth model movie (hereafter the \texttt{Freytag} model), we extracted a series of frames  from the intensity movie provided on the web site of B. Freytag.\footnote{\url{http://www.astro.uu.se/~bf/movie/intensity.html}} In total we selected a subset of 24 frames spanning a single (1.3~yr) stellar pulsation cycle to mimic a plausible monitoring schedule for the star of every few weeks. The original jpeg frames were translated into FITS files using the \texttt{ImageMagick} software\footnote{\url{https://imagemagick.org}} and further adapted for our needs as described above. The star was assumed to have a J2000 sky position of RA=02$^{\rm h}$ 19$^{\rm m}$, DEC=$-02^{\circ}$ 58$'$. 

\subsubsection{Model of an Evolving Red Supergiant\protect\label{Chiavassa}}
To simulate the time-varying appearance of the radio photosphere of an RSG star, we have adopted the $H$-band model st35gm03n07 from \cite{Chiavassa2009}. This model represents a 12$M_{\odot}$ RSG star with a bolometric luminosity $L=93,000L_{\odot}$, a mean effective temperature $T_{\rm eff}$=3490~K, and a radius $R=832R_{\odot}$. The resolution of the model is 8.6$R_{\odot}$. We adapt this model to represent a radio photosphere whose angular diameter and flux density at 46.1~GHz are $\sim$80~mas and 28~mJy, respectively, comparable to the RSG Betelgeuse, which lies at a distance of $\sim$222~pc \citep{Lim1998}. 

To formulate our ground truth model movie (hereafter the \texttt{Chiavassa} model), we extracted a series of 32 frames spanning $\sim$2~years from the intensity movie available on the web site of A. Chiavassa\footnote{\url{https://www-n.oca.eu/chiavassa/scarica/IONIC_rsun.mov}}.
The original jpeg frames were translated into FITS files and further adapted for our needs as described above. The star was assumed to have a J2000 sky position comparable to Betelgeuse (RA=05$^{\rm h}$ 55$^{\rm m}$, DEC=$+07^{\circ}$ 24$'$).

\section{Image Reconstructions \protect\label{imaging}}
\subsection{Multi-scale CLEAN\protect\label{MSCLEAN}}
For our \texttt{CLEAN} imaging tests, we used the \texttt{CASA} 5.4.0-70 version 
of multi-scale (\texttt{MS}) \texttt{CLEAN} as implemented via  the `\texttt{clean}' task. 
A general overview of \texttt{MS-CLEAN} can be found in e.g., \cite{Cornwell2008} \citep[see also][]{Rich2008}. For all \texttt{CLEAN} images presented in this work, we adopted uniform weighting, a loop gain of 0.1, a cell size of 0.2~mas, and used 25,000--50,000 \texttt{CLEAN}ing iterations, depending on the complexity of the model. No \texttt{CLEAN} boxes were used. We set the multi-scale parameter array in the \texttt{CLEAN} task to  [0,3,9,15,30,60,180,200,360] pixels for our uniform disk models (see Section~\ref{unidisk} below) and [0,3,9,15,30,60,180,300] pixels for both the \texttt{Freytag} and \texttt{Chiavassa} models (Sections~\ref{Freytag}, \ref{Chiavassa}). This combination of parameters was found to lead to generally good results for our model data sets. However, we did not attempt an exhaustive search of parameter space. For simplicity, we also made no attempt here to explore the effects of Briggs weighting \citep{Briggs99} and/or tapering on our resulting \texttt{MS-CLEAN} images. The effects of these parameters on ngVLA image quality have been investigated in previous studies by \cite{Carilli+2016, Carilli2016, Carilli2017, Carilli2018}; and \cite{Rosero2019}. 

\begin{figure*}[t]
\centering
\gridline{
    \fig{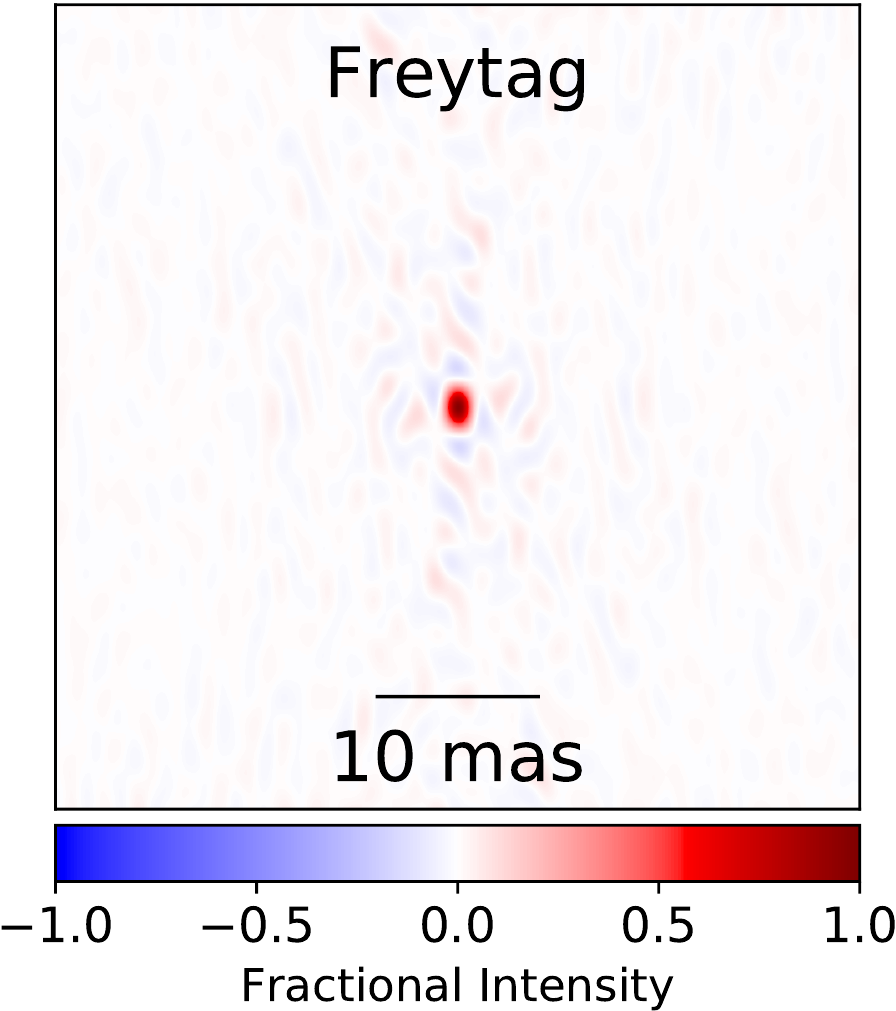}{0.30\textwidth}{(a) \texttt{Freytag} model}
    \fig{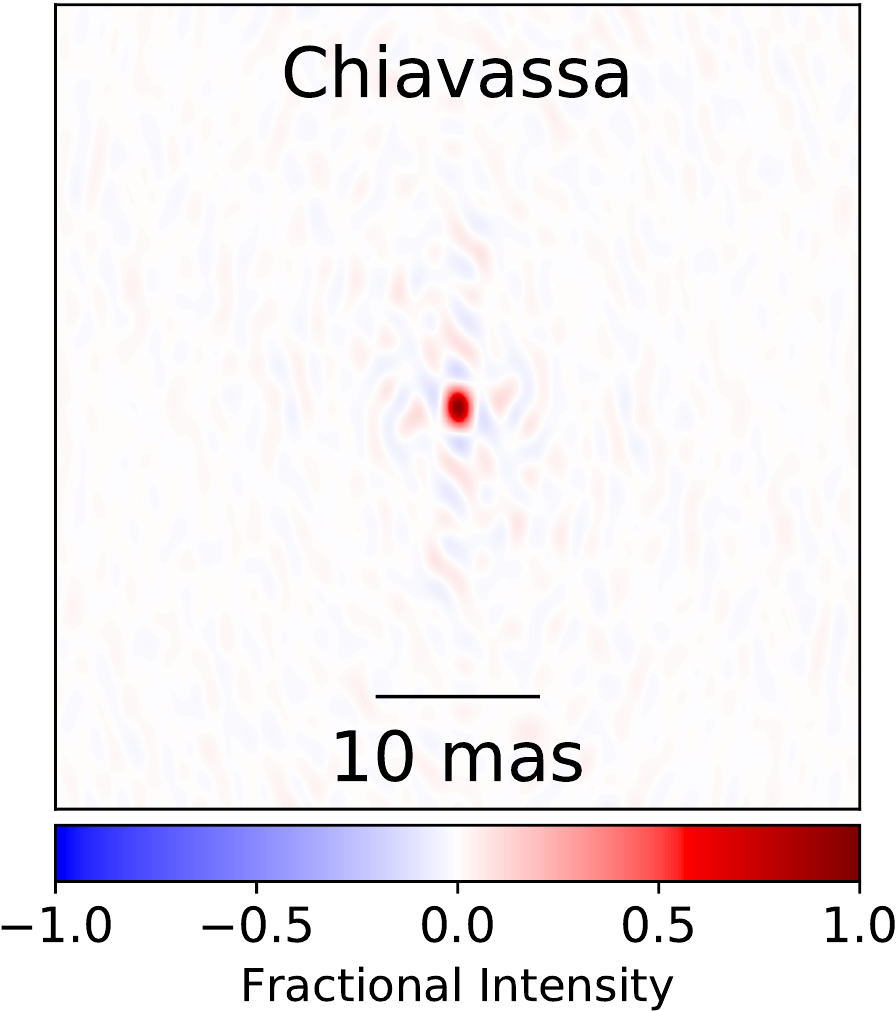}{0.30\textwidth}{(b) \texttt{Chiavassa} model}
    \fig{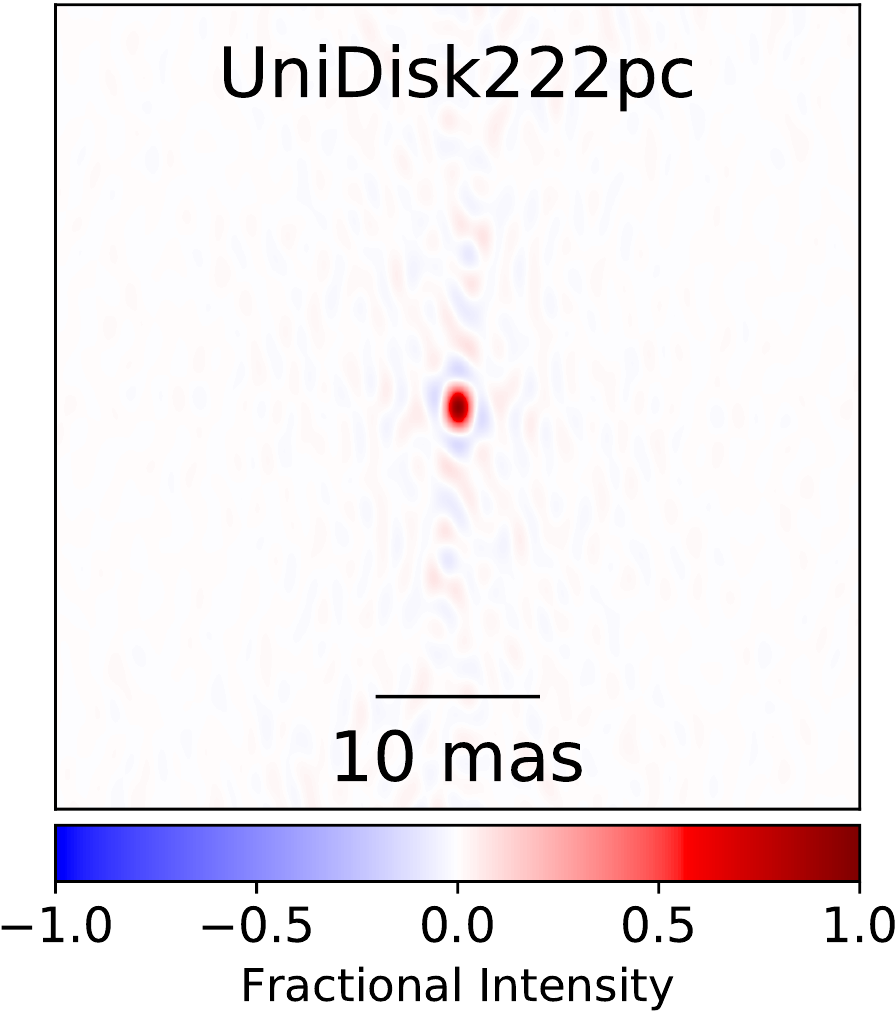}{0.30\textwidth}{(c) \texttt{UniDisk222pc} and \texttt{UniDisk1kpc} models}
}
\caption{Synthesized beams of simulated observations with uniform weighting. Table \ref{tab:beamprm} summarizes the FWHM size and position angle of each beam.
}
\label{fig:beam}
\end{figure*}

In Figure~\ref{fig:beam}, we show the uniform-weighted synthesized beam for each model that was used in the \texttt{MS-CLEAN} reconstruction. The corresponding beam parameters are summarized in Table \ref{tab:beamprm}. As shown in Figure \ref{fig:beam}, the synthesized beams do not exhibit a Gaussian-like decay from the beam center, but rather have linearly-scaled tails similar to Figure~\ref{fig:carilli_beam}, which are much more extended than the beam FWHM sizes and are particularly elongated along a roughly N-S direction.

\begin{table}[t]
    \centering
    \begin{tabular}{lccc}
        \hline \hline
        Model & $\theta _{\rm maj}$ & $\theta _{\rm min}$ & $\theta _{\rm PA}$\\
        & (mas) & (mas) & ($^{\circ}$)\\
        \hline
        \texttt{Freytag} & 2.1 & 1.3 & 1.3 \\
        \texttt{Chiavassa} & 1.9 & 1.3 & 3.6 \\
        \texttt{UniDisk} models & 2.0 & 1.2 & 0.8 \\ \hline
    \end{tabular}
    \caption{The parameters of the synthesized beams in Figure \ref{fig:beam} adopted for \texttt{MS-CLEAN} reconstructions.}
    \label{tab:beamprm}
\end{table}

\subsection{RML Imaging}
For our present RML imaging investigations we used \texttt{SMILI}\footnote{\url{https://github.com/astrosmili/smili}} \citep[][]{Akiyama2017a, Akiyama2017b} Version 0.1.0 \citep{smiliv010}, a python-interfaced open-source imaging library primarily developed for the EHT. Simulated data (see Section~\ref{modelsandsims}) were exported to uvfits files from \texttt{CASA} and loaded into \texttt{SMILI} for imaging and analysis. Since visibility weights in uvfits files from \texttt{CASA} do not reflect actual thermal noise, they were re-evaluated using the scatter in visibilities within 1~hour blocks using the \texttt{weightcal} method. Images are then reconstructed with full complex visibilities. 

The most relevant parameters for \texttt{SMILI} imaging (or more widely RML methods) are the pixel size and field-of-view of the image, and also the choice and weights of regularization functions. We adopt the pixel size of 0.2~mas for all of the models. The field-of-view is set to be 512 pixels for \texttt{Chiavassa} and \texttt{UniDisk222pc} models, 320 pixels for \texttt{Freytag} model, and 128 pixels for \texttt{UniDisk1kpc} model. 

For the uniform-disk models (\texttt{UniDisk222pc, UniDisk1kpc}) and \texttt{Chiavassa} model, we employ TV regularization \citep[see e.g.,][]{Rudin1992, Akiyama2017a, Akiyama2017b}. Images were reconstructed for regularization parameters of $[10^0,\,10^1,...,10^5]$. Then for the final image the largest parameter was adopted that gave a reduced $\chi$-square close to unity and also residuals consistent with the normal distribution. The selected parameters were $10^4$, $10^2$, and $10^4$, respectively. For the \texttt{Freytag} model, we employ a relative entropy term \citep[e.g., see][]{EHTC2019d} with a flat prior for the reconstruction. Images were reconstructed for regularization parameters of $[10^{-4},\,10^{-3},...,10^2]$, and the parameter of $10^{-2}$ was selected in the same manner as for the other models. 

\section{Results\protect\label{results}}
\begin{figure*}[t]
\centering
\gridline{
    \fig{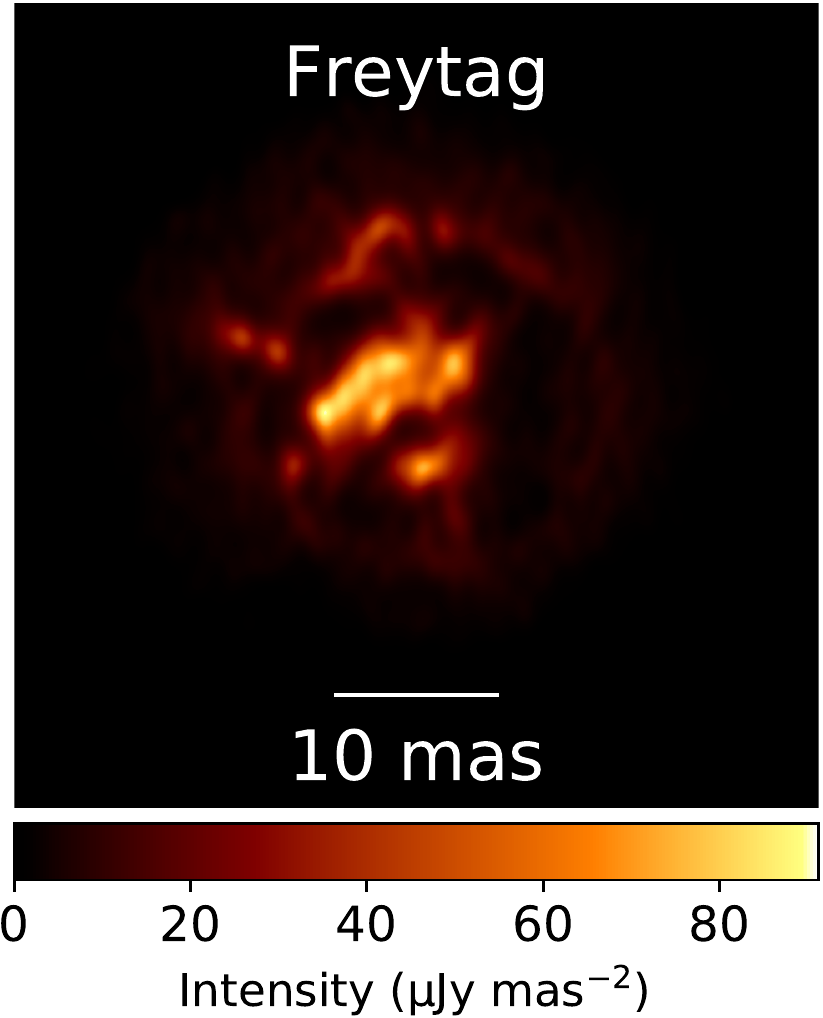}{0.25\textwidth}{(a) \texttt{Freytag}}
    \fig{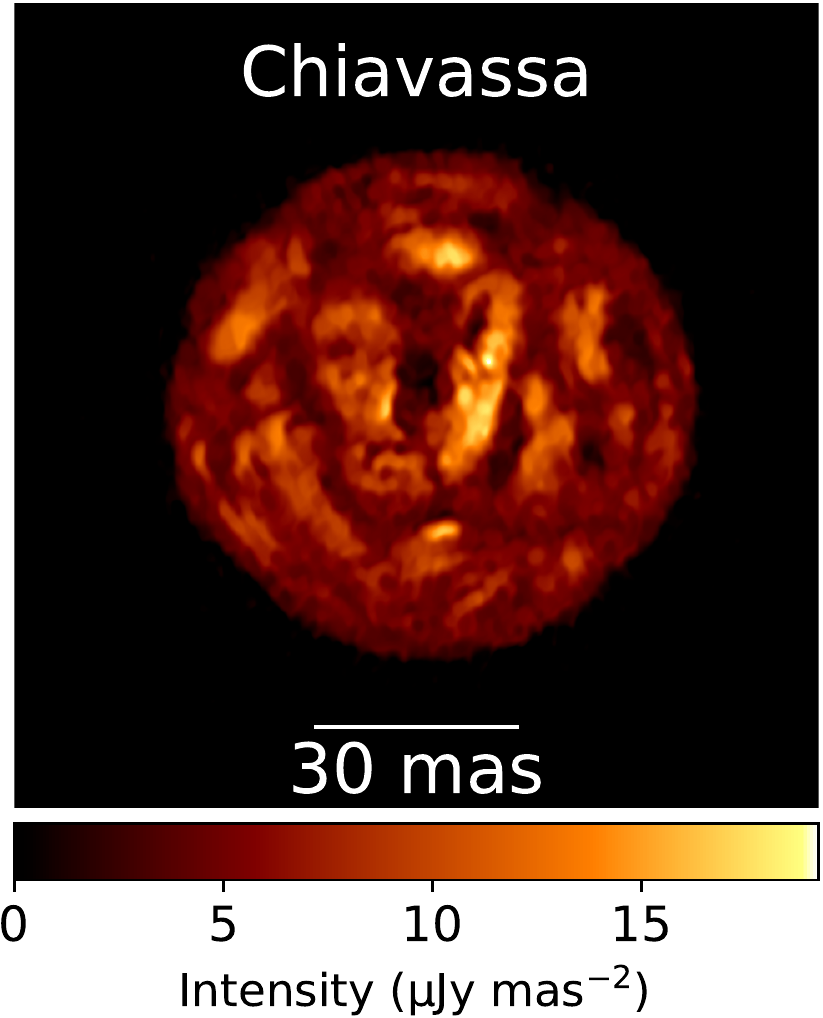}{0.25\textwidth}{(b) \texttt{Chiavassa}}
    \fig{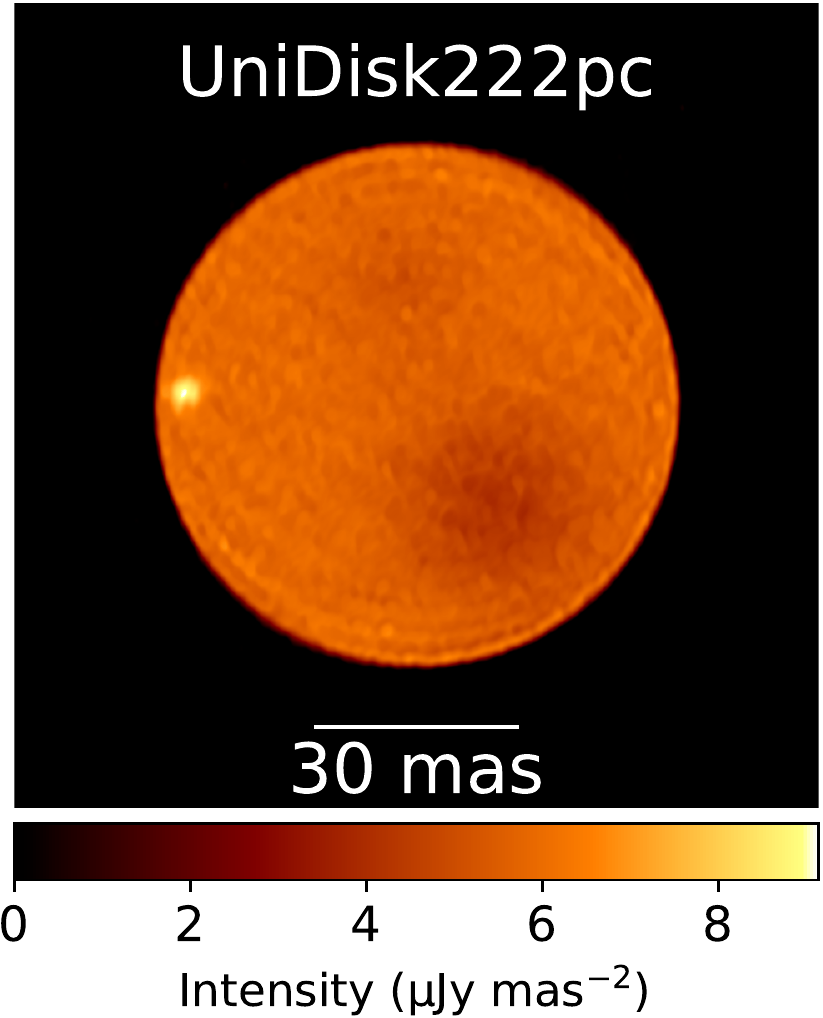}{0.25\textwidth}{(c) \texttt{UniDisk222pc}}
    \fig{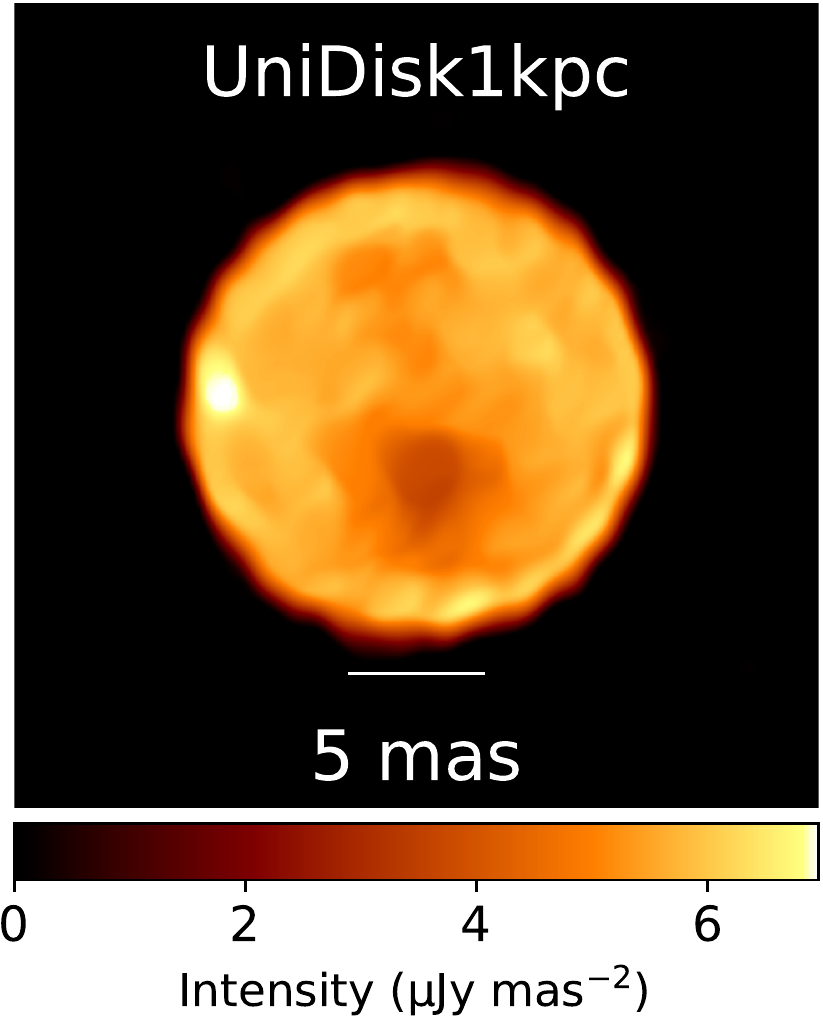}{0.25\textwidth}{(b) \texttt{UniDisk1kpc}}
}
\caption{
\texttt{SMILI} reconstructions of all four stellar models without any beam convolution. For the \texttt{Freytag} and \texttt{Chiavassa} models, only the first frame of the full time sequence is shown.
}
\label{fig:smiliimages}
\end{figure*}
\begin{figure*}[t]
\centering
\gridline{
    \fig{{images/FRAME001.resid.scale1.00}.pdf}{0.8\textwidth}{\texttt{Freytag} -- Scale: 1.0}
}
\gridline{
    \fig{{images/FRAME001.resid.scale0.50}.pdf}{0.8\textwidth}{\texttt{Freytag} -- Scale: 0.5}
}
\caption{
\texttt{SMILI} and \texttt{CASA} reconstructions and their residuals for all four stellar models. For the \texttt{Freytag} and \texttt{Chiavassa} models, only the first frame of the full time sequence is shown. In each row, we show the ground truth image, reconstructed image, and residual image. To illustrate the fidelity at the nominal \texttt{CLEAN} resolution, the top panels are convolved with the elliptical Gaussian beam used for uniform weighting in the \texttt{CASA} (Section~\ref{MSCLEAN}) imaging (scale=1.0). The lower panels are convolved with a beam half that size (scale=0.5) to show the effects of mild super-resolution. The FWHM size of the convolving beam is shown by the ellipse on each panel (see also Table~1). (continued to the next page.) 
}
\label{fig:smilivscasa}
\end{figure*}
\addtocounter{figure}{-1}
\begin{figure*}[!ht]
\centering 
\gridline{
    \fig{{images/Chiavassa.resid.scale1.00}.pdf}{0.8\textwidth}{\texttt{Chiavassa} -- Scale: 1.0}
}
\gridline{
    \fig{{images/Chiavassa.resid.scale0.50}.pdf}{0.8\textwidth}{\texttt{Chiavassa} -- Scale: 0.5}
}
\caption{--- continued.}
\end{figure*}
\addtocounter{figure}{-1}
\begin{figure*}[!ht]
\centering 
\gridline{
    \fig{{images/UniDisk222pc.resid.scale1.00}.pdf}{0.8\textwidth}{\texttt{UniDisk222pc} -- Scale: 1.0}
}
\gridline{
    \fig{{images/UniDisk222pc.resid.scale0.50}.pdf}{0.8\textwidth}{\texttt{UniDisk222pc} -- Scale: 0.5}
}
\caption{--- continued.}
\end{figure*}
\addtocounter{figure}{-1}
\begin{figure*}[!ht]
\centering 
\gridline{
    \fig{{images/UniDisk1kpc.resid.scale1.00}.pdf}{0.8\textwidth}{\texttt{UniDisk1kpc} -- Scale: 1.0}
}
\gridline{
    \fig{{images/UniDisk1kpc.resid.scale0.50}.pdf}{0.8\textwidth}{\texttt{UniDisk1kpc} -- Scale: 0.5}
}
\caption{--- continued.}
\end{figure*}

\begin{figure*}[t]
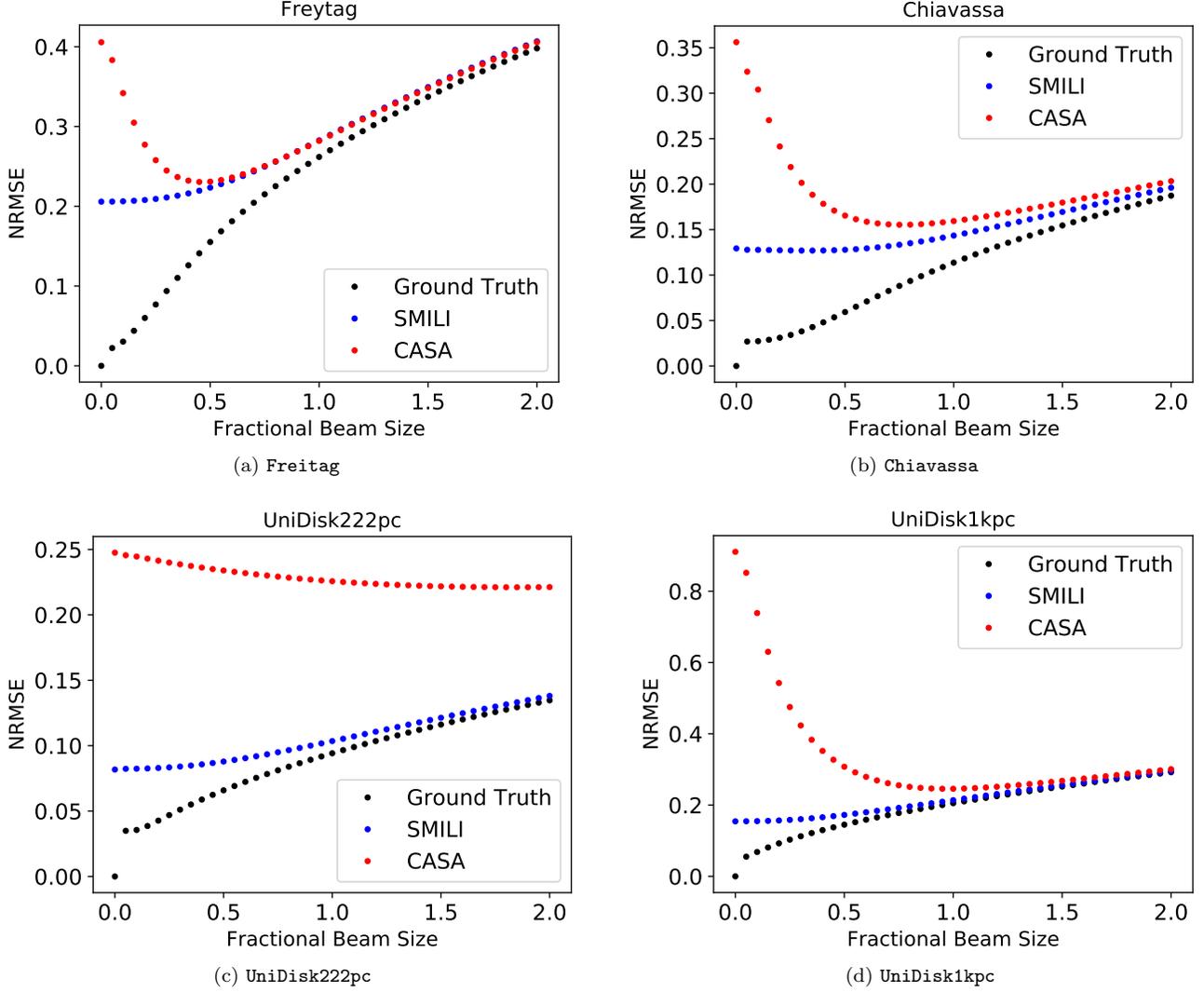

\centering 
\gridline{
    \fig{{images/FRAME001.nrmse_profile}.pdf}{0.44\textwidth}{(a) \texttt{Freitag}}
    \fig{{images/Chiavassa.nrmse_profile}.pdf}{0.45\textwidth}{(b) \texttt{Chiavassa}}
}
\gridline{
    \fig{{images/UniDisk222pc.nrmse_profile}.pdf}{0.45\textwidth}{(c) \texttt{UniDisk222pc}}
    \fig{{images/UniDisk1kpc.nrmse_profile}.pdf}{0.44\textwidth}{(d) \texttt{UniDisk1kpc}}
}
\caption{The normalized root-mean-square errors (NRMSEs) of reconstructions as a function of the restoring beam size. Each NRMSE curve was calculated between the corresponding beam-convolved image and the {\it non-convolved} ground truth image adopted as the reference. The beam size on the horizontal axis is normalized to that of uniform weighting used in \texttt{CASA} imaging.}
\label{fig:nrmse}
\end{figure*}

In Figure \ref{fig:smiliimages}, we show RML reconstructions with \texttt{SMILI}, which are not beam-convolved. \texttt{SMILI} can reconstruct piecewise smooth images consistent with given data sets even without the restoring beam, thanks to various regularization functions. 

We show more detailed comparisons with the ground truth and \texttt{CASA} \texttt{MS-CLEAN} images at the resolution of uniform weighting and also two times finer resolution in Figure \ref{fig:smilivscasa}. Both the \texttt{CASA} and \texttt{SMILI} images reconstruct most of representative features in the ground truth images,  underscoring the ngVLA's capability for imaging stellar photospheres in exquisite detail. The RMS noise in the \texttt{MS-CLEAN} images estimated by the background regions are 1.67~$\rm \mu $Jy/beam, 1.9~$\rm \mu $Jy/beam, 2.9~$\rm \mu $Jy/beam and 1.0~$\rm \mu $Jy/beam, providing the dynamic range to the peak intensity of $\sim$140, 26, 8 and 18 for the \texttt{Freytag}, \texttt{Chiavassa}, \texttt{UniDisk222pc} and \texttt{UniDisk1kpc} models, respectively. Residual errors seen in Figure \ref{fig:smilivscasa} are greater than these RMS noise levels\footnote{We do not show the traditional RMS noise and dynamic range estimated from the residual maps for \texttt{SMILI}, since \texttt{SMILI} does not use the dirty map, the dirty beam, or even $uv-$gridding. \texttt{SMILI} reconstructions are equivalent to imaging with natural weighting and also provide better fits to the data (see Table \ref{tab:chisq}), which should result in less RMS noise and higher dynamic range and indicate residuals that are not dominated by thermal errors.}, 
suggesting that they are not sensitivity limited (i.e. thermal-error dominated) but dynamic-range limited, where the image fidelity is predominantly limited by both the $uv$-coverage of the observations and the performance of the imaging algorithms.  

\texttt{SMILI} provides better image quality for all four models. In particular, differences in the quality of feature reconstructions are clear for the uniform disk models; \texttt{SMILI} successfully reconstructs both brighter and fainter spots, while some of them do not clearly appear in \texttt{CASA} images. Another obvious advantage of RML methods is seen in the localization of emission ---\texttt{SMILI} locates much less emission outside of the stellar photosphere than \texttt{CASA} \texttt{MS-CLEAN}, although the use of \texttt{CLEAN} boxes may help to mitigate this effect. The \texttt{MS-CLEAN} images seem to have a noise floor at a level of $\lesssim 10$~\% of the peak intensity spread in the image field of view, which seems comparable with the typical level of side lobes in the synthesized beam (see Figure \ref{fig:carilli_beam} and \ref{fig:beam}). This indicates that \texttt{CASA} \texttt{MS-CLEAN} needs a careful handling of $uv-$weightings to minimize the effects of sidelobes as discussed, for instance, in \citet{Carilli+2018}.

For the photosphere emission, \texttt{SMILI} images have residuals of $\lesssim 10$~\% better than those of \texttt{CASA} at the full resolution of uniform weighting and even at half of its beam size ($\sim 0.3 \lambda / D$). Considering that the typical beam size with uniform weighting is $\sim 0.6 \lambda /D$, where $\lambda$ is the observing wavelength and $D$ is the maximum baseline length, this demonstrates that RML indeed may improve the fidelity of ngVLA images at high angular resolution, up to resolutions modestly finer than the diffraction limit $\lambda / D$. 

For a more quantitative comparison on multiple scales, in Figure \ref{fig:nrmse} we show characteristic levels of reconstruction errors at each spatial scale using the normalized root-mean-square error \citep[NRMSE;][]{Chael2016}. NRMSE is defined by
\begin{equation}
    {\rm NRMSE}({\bf I},\,{\bf K}) = \sqrt{\frac{\sum_i |I_i - K_i|^2}{\sum_i |K_i|^2}},
\end{equation}
where ${\bf I}$ is the image to be evaluated and ${\bf K}$ is the reference image. We adopt the non-convolved ground truth image as the reference image, and evaluate NRMSEs of the ground truth and reconstructed images convolved with an elliptical Gaussian beam equivalent to the one appropriate for uniform weighting. The curve for the ground truth image shows the loss in the image fidelity due to the limited angular resolution. Except for the {\tt Freytag} model with its many compact emission features, RML reconstructions with {\tt SMILI} outperform \texttt{MS-CLEAN} reconstructions with {\tt CASA} for a wide range of spatial scales including the nominal resolution at uniform weighting. 

\texttt{SMILI} also shows better goodness-of-fit than \texttt{CASA} for all four models. In Table \ref{tab:chisq}, we show the mean $\chi ^2$ values (i.e. similar to the reduced $\chi ^2$ value for deterministic problems) of each reconstruction. RML reconstructions with \texttt{SMILI} enable derivation of images well consistent with the data sets for given thermal error budgets, while \texttt{CASA} shows larger $\chi ^2$, presumably attributed to difficulties of convergence to an optimal solution. In particular, the convergence issue severely affects the \texttt{MS-CLEAN} fits to \texttt{UniDisk222pc}, which has the most uniform and extended emission.

\begin{table}[t]
    \centering
    \begin{tabular}{lcc}
        \hline \hline
        Model & \texttt{SMILI} & \texttt{CASA}\\ \hline
        \texttt{Freytag} & 1.00 & 1.05\\
        \texttt{Chiavassa} & 1.00 & 1.05\\
        \texttt{UniDisk222pc} & 1.00 & 32.70\\
        \texttt{UniDisk1kpc} & 1.00 & 1.01 \\\hline
    \end{tabular}
    \caption{Mean $\chi ^2$ for the full complex visibilities of the  reconstructions. Here, errors on the data are rescaled such that the ground truth images provide a mean $\chi ^2$ of unity.}
    \label{tab:chisq}
\end{table}

\subsection{Simulated Movies\protect\label{movies}}
As described above, an intriguing and groundbreaking science case for observing evolved stars with the ngVLA will be capturing the dynamic and complex kinematics of their stellar surfaces with multi-epoch imaging. For example, AGB stars such as Mira variables undergo regular radial pulsations of periods of order 1 year during which their visual brightness can change by a factor of $\sim$1000
\citep{Reid2002}. The radius and brightness temperature of the radio photosphere are also predicted to vary measurably over this time interval.\footnote{The \texttt{Freytag} AGB star model adopted here shows flux variations of $\sim$10\,\%. This is much less extreme than observed in some of the most highly time-variable AGB stars at visible wavelengths, but is comparable to variations seen in radio photospheres at cm wavelengths \citep{RM97}.} In addition, features such as giant convective cells on the surfaces of AGB stars are expected to evolve on timescales ranging from weeks to years owing to the complex interplay between pulsation, shocks, and convection \citep[e.g.,][]{Freytag2008, Freytag2017}. All of these effects are expected to lead to observable month-to-month changes in the properties of radio photospheres over the course of a pulsation cycle that will become readily observable at radio wavelengths for the first time with ngVLA. Here we have made a first attempt to emulate this by creating simulated movies of time-varying stellar emission observed with the ngVLA.

Figure~\ref{fig:smiliimages} and \ref{fig:smilivscasa} show only a single time frame from our \texttt{Freytag} and \texttt{Chiavassa} models for illustrative purposes. However, as noted in Sections~\ref{Freytag} and \ref{Chiavassa}, in both cases we have imaged a sequence of multiple frames, providing simulated ``movies'' of how the appearance of the stars may evolve over timescales of weeks to months. The full movies are available at the web site\footnote{\url{http://library.nrao.edu/ngvla66sppl.shtml}} of National Radio Astronomy Observatory.

\section{Discussion and Future Prospects\protect\label{discussion}}
In the present work, we have demonstrated that the ngVLA is capable of well resolving the surfaces of nearby stars, which are currently only marginally resolved with the existing interferometers such as the VLA and ALMA. Furthermore, with {\tt SMILI}, we have shown that the state-of-the-art RML imaging techniques may provide further improvements in the image fidelity and capture scientifically meaningful features more accurately than \texttt{MS-CLEAN} reconstructions. Here, we outline possibilities for future studies.

First, the current simulations only handle thermal noise assuming that data are calibrated accurately. However, in more realistic situations, we expect residual calibration errors in both the amplitudes and phases of the complex visibilities, especially, on longer baselines (reaching milliarcsecond resolutions) where calibrators are often no longer point sources. Indeed, RML methods, which can include error budgets for systematic errors \citep{EHTC2019d} or directly use closure quantities free from station-based gain errors \citep[e.g.][]{Chael2016, Bouman2016, Akiyama2017a, Chael2018}, generally provide better reconstructions than \texttt{CLEAN}  for VLBI imaging where systematic errors tend to be large \citep{EHTC2019d}. As a next step we will test both imaging techniques on ngVLA simulations that include more realistic calibration errors. At this stage, we will also need to explore a wider range of parameters for \texttt{MS-CLEAN} than in the present work. 

Spectral line imaging (effectively adding an extra dimension to the continuum imaging presented in this work) is another intriguing application that should be studied. Numerous astrophysically interesting spectral lines will fall in the cm and mm bands covered by the ngVLA \citep[e.g.,][]{Murphy2018}. For example, the cool, extended
atmospheres and circumstellar environments of AGB and RSG stars
give rise to rotational transitions from a multitude of molecules which can be used to probe chemistry, temperature, and density, as well as
wind outflow speeds and atmospheric kinematics \citep{MattClaussen2018}. Building from the continuum case explored here, RML reconstructions may be used to improve ngVLA spectral line imaging by simply applying them on a channel-by-channel basis. Furthermore, recent developments of dynamical imaging \citep{Johnson2017,Bouman2018} demonstrate that the fidelity of three-dimensional imaging can be significantly improved by simultaneously reconstructing all images with additional regularization functions leading to piece-wise smooth variations on the third dimension (i.e. frequency). However, the performance of RML reconstructions on spectral line imaging has not been tested in the past literature, and therefore is an important topic for the future work.

\clearpage
\section*{Acknowledgments} \noindent
We thank Eric Murphy for fruitful discussions and many useful suggestions for this work. 
We also thank Bernd Freytag for granting permission to use results from his 3D stellar models and for useful comments on this memo. We are grateful to Andrea Isella and Luca Ricci for discussions that helped to motivate this study.
This work was supported by the ngVLA Community Studies program, coordinated by the National Radio Astronomy Observatory (NRAO), which is a facility of the National Science Foundation (NSF) operated under cooperative agreement by Associated Universities, Inc.
K.A. is a Jansky fellow of the NRAO. Developments of \texttt{SMILI} at MIT Haystack Observatory have been financially supported by grants from the NSF (AST-1440254; AST-1614868).
The Black Hole Initiative at Harvard University is financially supported by a grant from the John Templeton Foundation. 

\bibliography{bibtex_kazu}{}

\begin{thebibliography}{}
\providecommand\natexlab[1]{#1}
\providecommand\JournalTitle[1]{#1}

\bibitem[{Akiyama {et~al.}(2019)Akiyama, Moriyama, Cho, Tazaki, Ikeda, Sasada,
  Okino, \& Honma}]{smiliv010}
Akiyama, K., Moriyama, K., Cho, I., {et~al.} 2019,
  \href{http://dx.doi.org/10.5281/zenodo.3459837}{\JournalTitle{Zenodo},
  3459837}

\bibitem[{{Akiyama} {et~al.}(2017{\natexlab{a}}){Akiyama}, {Kuramochi},
  {Ikeda}, {Fish}, {Tazaki}, {Honma}, {Doeleman}, {Broderick}, {Dexter},
  {Mo{\'s}cibrodzka}, {Bouman}, {Chael}, \& {Zaizen}}]{Akiyama2017a}
{Akiyama}, K., {Kuramochi}, K., {Ikeda}, S., {et~al.} 2017{\natexlab{a}},
  \href{http://dx.doi.org/10.3847/1538-4357/aa6305}{\JournalTitle{\apj}, 838,
  1}

\bibitem[{{Akiyama} {et~al.}(2017{\natexlab{b}}){Akiyama}, {Ikeda}, {Pleau},
  {Fish}, {Tazaki}, {Kuramochi}, {Broderick}, {Dexter}, {Mo{\'s}cibrodzka},
  {Gowanlock}, {Honma}, \& {Doeleman}}]{Akiyama2017b}
{Akiyama}, K., {Ikeda}, S., {Pleau}, M., {et~al.} 2017{\natexlab{b}},
  \href{http://dx.doi.org/10.3847/1538-3881/aa6302}{\JournalTitle{\aj}, 153,
  159}

\bibitem[{{Bouman} {et~al.}(2018){Bouman}, {Johnson}, {Dalca}, {Chael},
  {Roelofs}, {Doeleman}, \& {Freeman}}]{Bouman2018}
{Bouman}, K.~L., {Johnson}, M.~D., {Dalca}, A.~V., {et~al.} 2018,
  \href{http://dx.doi.org/10.1109/TCI.2018.2838452}{\JournalTitle{IEEE
  Transactions on Computational Imaging}, 4, 512}

\bibitem[{{Bouman} {et~al.}(2016){Bouman}, {Johnson}, {Zoran}, {Fish},
  {Doeleman}, \& {Freeman}}]{Bouman2016}
{Bouman}, K.~L., {Johnson}, M.~D., {Zoran}, D., {et~al.} 2016, in The IEEE
  Conference on Computer Vision and Pattern Recognition (CVPR), 913

\bibitem[{{Briggs} {et~al.}(1999){Briggs}, {Schwab}, \& {Sramek}}]{Briggs99}
{Briggs}, D.~S., {Schwab}, F.~R., \& {Sramek}, R.~A. 1999, in Astronomical
  Society of the Pacific Conference Series, Vol. 180, Synthesis Imaging in
  Radio Astronomy II, ed. G.~B. {Taylor}, C.~L. {Carilli}, \& R.~A. {Perley},
  127

\bibitem[{{Carilli}(2016)}]{Carilli2016}
{Carilli}, C.~L. 2016, ngVLA Memo No. 12

\bibitem[{{Carilli}(2017)}]{Carilli2017}
---. 2017, {ngVLA Memo No. 16}

\bibitem[{{Carilli}(2018)}]{Carilli2018}
---. 2018, ngVLA Memo No. 47

\bibitem[{{Carilli} {et~al.}(2018{\natexlab{a}}){Carilli}, {Butler}, {Golap},
  {Carilli}, \& {White}}]{Carilli2018ngvla}
{Carilli}, C.~L., {Butler}, B., {Golap}, K., {Carilli}, M.~T., \& {White},
  S.~M. 2018{\natexlab{a}}, in Astronomical Society of the Pacific Conference
  Series, Vol. 517, Science with a Next Generation Very Large Array, ed.
  E.~{Murphy}, 369

\bibitem[{{Carilli} {et~al.}(2018{\natexlab{b}}){Carilli}, {Erickson},
  {Greisen}, \& {the ngVLA Team}}]{Carilli+2018}
{Carilli}, C.~L., {Erickson}, E., {Greisen}, E., \& {the ngVLA Team}.
  2018{\natexlab{b}}, The Next Generation Very Large Array: Configuration,
  \url{https://ngvla.nrao.edu/download/MediaFile/91/original}

\bibitem[{{Carilli} {et~al.}(2017){Carilli}, {Greisen}, {Nyland}, \&
  {Indebetouw}}]{Carilli+2017a}
{Carilli}, C.~L., {Greisen}, E., {Nyland}, K., \& {Indebetouw}, R. 2017,
  Instructions for using CASA simulator for the ngVLA

\bibitem[{{Carilli} {et~al.}(2016){Carilli}, {Ricci}, {Barge}, \&
  {Clark}}]{Carilli+2016}
{Carilli}, C.~L., {Ricci}, L., {Barge}, P., \& {Clark}, B. 2016, ngVLA Memo No.
  11

\bibitem[{{Chael} {et~al.}(2018){Chael}, {Johnson}, {Bouman}, {Blackburn},
  {Akiyama}, \& {Narayan}}]{Chael2018}
{Chael}, A.~A., {Johnson}, M.~D., {Bouman}, K.~L., {et~al.} 2018,
  \href{http://dx.doi.org/10.3847/1538-4357/aab6a8}{\JournalTitle{\apj}, 857,
  23}

\bibitem[{{Chael} {et~al.}(2016){Chael}, {Johnson}, {Narayan}, {Doeleman},
  {Wardle}, \& {Bouman}}]{Chael2016}
{Chael}, A.~A., {Johnson}, M.~D., {Narayan}, R., {et~al.} 2016,
  \href{http://dx.doi.org/10.3847/0004-637X/829/1/11}{\JournalTitle{\apj}, 829,
  11}

\bibitem[{{Chiavassa} {et~al.}(2009){Chiavassa}, {Plez}, {Josselin}, \&
  {Freytag}}]{Chiavassa2009}
{Chiavassa}, A., {Plez}, B., {Josselin}, E., \& {Freytag}, B. 2009,
  \href{http://dx.doi.org/10.1051/0004-6361/200911780}{\JournalTitle{\aap},
  506, 1351}

\bibitem[{{Cornwell}(2008)}]{Cornwell2008}
{Cornwell}, T.~J. 2008,
  \href{http://dx.doi.org/10.1109/JSTSP.2008.2006388}{\JournalTitle{IEEE
  Journal of Selected Topics in Signal Processing}, 2, 793}

\bibitem[{{Event Horizon Telescope
  Collaboration}(2019{\natexlab{a}})}]{EHTC2019d}
{Event Horizon Telescope Collaboration}. 2019{\natexlab{a}},
  \href{http://dx.doi.org/10.3847/2041-8213/ab0e85}{\JournalTitle{\apjl}, 875,
  L4}

\bibitem[{{Event Horizon Telescope
  Collaboration}(2019{\natexlab{b}})}]{EHTC2019b}
---. 2019{\natexlab{b}},
  \href{http://dx.doi.org/10.3847/2041-8213/ab0e85}{\JournalTitle{\apjl}, 875,
  L2}

\bibitem[{{Fish} {et~al.}(2016){Fish}, {Akiyama}, {Bouman}, {Chael}, {Johnson},
  {Doeleman}, {Blackburn}, {Wardle}, \& {Freeman}}]{Fish2016}
{Fish}, V., {Akiyama}, K., {Bouman}, K., {et~al.} 2016,
  \href{http://dx.doi.org/10.3390/galaxies4040054}{\JournalTitle{Galaxies}, 4,
  54}

\bibitem[{{Freytag} \& {H{\"o}fner}(2008)}]{Freytag2008}
{Freytag}, B., \& {H{\"o}fner}, S. 2008,
  \href{http://dx.doi.org/10.1051/0004-6361:20078096}{\JournalTitle{\aap}, 483,
  571}

\bibitem[{{Freytag} {et~al.}(2017){Freytag}, {Liljegren}, \&
  {H{\"o}fner}}]{Freytag2017}
{Freytag}, B., {Liljegren}, S., \& {H{\"o}fner}, S. 2017,
  \href{http://dx.doi.org/10.1051/0004-6361/201629594}{\JournalTitle{\aap},
  600, A137}

\bibitem[{{Harper}(2018)}]{Harper2018}
{Harper}, G.~M. 2018, in Astronomical Society of the Pacific Conference Series,
  Vol. 517, Science with a Next Generation Very Large Array, ed. E.~{Murphy},
  265

\bibitem[{{H{\"o}gbom}(1974)}]{Hogbom74}
{H{\"o}gbom}, J.~A. 1974, \JournalTitle{\aaps}, 15, 417

\bibitem[{{Honma} {et~al.}(2014){Honma}, {Akiyama}, {Uemura}, \&
  {Ikeda}}]{Honma2014}
{Honma}, M., {Akiyama}, K., {Uemura}, M., \& {Ikeda}, S. 2014,
  \href{http://dx.doi.org/10.1093/pasj/psu070}{\JournalTitle{\pasj}, 66, 95}

\bibitem[{{Ikeda} {et~al.}(2016){Ikeda}, {Tazaki}, {Akiyama}, {Hada}, \&
  {Honma}}]{Ikeda2016}
{Ikeda}, S., {Tazaki}, F., {Akiyama}, K., {Hada}, K., \& {Honma}, M. 2016,
  \href{http://dx.doi.org/10.1093/pasj/psw042}{\JournalTitle{\pasj}, 68, 45}

\bibitem[{{Johnson} {et~al.}(2017){Johnson}, {Bouman}, {Blackburn}, {Chael},
  {Rosen}, {Shiokawa}, {Roelofs}, {Akiyama}, {Fish}, \&
  {Doeleman}}]{Johnson2017}
{Johnson}, M.~D., {Bouman}, K.~L., {Blackburn}, L., {et~al.} 2017,
  \href{http://dx.doi.org/10.3847/1538-4357/aa97dd}{\JournalTitle{\apj}, 850,
  172}

\bibitem[{{Kuramochi} {et~al.}(2018){Kuramochi}, {Akiyama}, {Ikeda}, {Tazaki},
  {Fish}, {Pu}, {Asada}, \& {Honma}}]{Kuramochi2018}
{Kuramochi}, K., {Akiyama}, K., {Ikeda}, S., {et~al.} 2018,
  \href{http://dx.doi.org/10.3847/1538-4357/aab6b5}{\JournalTitle{\apj}, 858,
  56}

\bibitem[{{Liljegren} {et~al.}(2018){Liljegren}, {H{\"o}fner}, {Freytag}, \&
  {Bladh}}]{Liljegren2018}
{Liljegren}, S., {H{\"o}fner}, S., {Freytag}, B., \& {Bladh}, S. 2018,
  \href{http://dx.doi.org/10.1051/0004-6361/201833203}{\JournalTitle{\aap},
  619, A47}

\bibitem[{{Lim} {et~al.}(1998){Lim}, {Carilli}, {White}, {Beasley}, \&
  {Marson}}]{Lim1998}
{Lim}, J., {Carilli}, C.~L., {White}, S.~M., {Beasley}, A.~J., \& {Marson},
  R.~G. 1998, \href{http://dx.doi.org/10.1038/33352}{\JournalTitle{\nat}, 392,
  575}

\bibitem[{{Lu} {et~al.}(2014){Lu}, {Broderick}, {Baron}, {Monnier}, {Fish},
  {Doeleman}, \& {Pankratius}}]{Lu2014}
{Lu}, R.-S., {Broderick}, A.~E., {Baron}, F., {et~al.} 2014,
  \href{http://dx.doi.org/10.1088/0004-637X/788/2/120}{\JournalTitle{\apj},
  788, 120}

\bibitem[{{Matthews} \& {Claussen}(2018)}]{MattClaussen2018}
{Matthews}, L.~D., \& {Claussen}, M.~J. 2018, in Astronomical Society of the
  Pacific Conference Series, Vol. 517, Science with a Next Generation Very
  Large Array, ed. E.~{Murphy}, 281

\bibitem[{{Matthews} {et~al.}(2015){Matthews}, {Reid}, \&
  {Menten}}]{Matthews2015}
{Matthews}, L.~D., {Reid}, M.~J., \& {Menten}, K.~M. 2015,
  \href{http://dx.doi.org/10.1088/0004-637X/808/1/36}{\JournalTitle{\apj}, 808,
  36}

\bibitem[{{Matthews} {et~al.}(2018){Matthews}, {Reid}, {Menten}, \&
  {Akiyama}}]{Matt+2018}
{Matthews}, L.~D., {Reid}, M.~J., {Menten}, K.~M., \& {Akiyama}, K. 2018,
  \href{http://dx.doi.org/10.3847/1538-3881/aac491}{\JournalTitle{\aj}, 156,
  15}

\bibitem[{{Menten} {et~al.}(2012){Menten}, {Reid}, {Kami{\'n}ski}, \&
  {Claussen}}]{Menten2012}
{Menten}, K.~M., {Reid}, M.~J., {Kami{\'n}ski}, T., \& {Claussen}, M.~J. 2012,
  \href{http://dx.doi.org/10.1051/0004-6361/201219422}{\JournalTitle{\aap},
  543, A73}

\bibitem[{{Murphy} {et~al.}(2018){Murphy}, {Bolatto}, {Chatterjee}, {Casey},
  {Chomiuk}, {Dale}, {de Pater}, {Dickinson}, {Francesco}, {Hallinan},
  {Isella}, {Kohno}, {Kulkarni}, {Lang}, {Lazio}, {Leroy}, {Loinard},
  {Maccarone}, {Matthews}, {Osten}, {Reid}, {Riechers}, {Sakai}, {Walter}, \&
  {Wilner}}]{Murphy2018}
{Murphy}, E.~J., {Bolatto}, A., {Chatterjee}, S., {et~al.} 2018, in
  Astronomical Society of the Pacific Conference Series, Vol. 517, Science with
  a Next Generation Very Large Array, ed. E.~{Murphy}, 3

\bibitem[{{O'Gorman} {et~al.}(2015){O'Gorman}, {Harper}, {Brown}, {Guinan},
  {Richards}, {Vlemmings}, \& {Wasatonic}}]{OGorman2015}
{O'Gorman}, E., {Harper}, G.~M., {Brown}, A., {et~al.} 2015,
  \href{http://dx.doi.org/10.1051/0004-6361/201526136}{\JournalTitle{\aap},
  580, A101}

\bibitem[{{Reid} \& {Goldston}(2002)}]{Reid2002}
{Reid}, M.~J., \& {Goldston}, J.~E. 2002,
  \href{http://dx.doi.org/10.1086/338947}{\JournalTitle{\apj}, 568, 931}

\bibitem[{{Reid} \& {Menten}(1997)}]{RM97}
{Reid}, M.~J., \& {Menten}, K.~M. 1997,
  \href{http://dx.doi.org/10.1086/303614}{\JournalTitle{\apj}, 476, 327}

\bibitem[{{Reid} \& {Menten}(2007)}]{RM07}
---. 2007, \href{http://dx.doi.org/10.1086/523085}{\JournalTitle{\apj}, 671,
  2068}

\bibitem[{{Rich} {et~al.}(2008){Rich}, {de Blok}, {Cornwell}, {Brinks},
  {Walter}, {Bagetakos}, \& {Kennicutt}}]{Rich2008}
{Rich}, J.~W., {de Blok}, W.~J.~G., {Cornwell}, T.~J., {et~al.} 2008,
  \href{http://dx.doi.org/10.1088/0004-6256/136/6/2897}{\JournalTitle{\aj},
  136, 2897}

\bibitem[{{Rosero}(2019)}]{Rosero2019}
{Rosero}, V. 2019, ngVLA Memo No. 55

\bibitem[{{Rudin} {et~al.}(1992){Rudin}, {Osher}, \& {Fatemi}}]{Rudin1992}
{Rudin}, L.~I., {Osher}, S., \& {Fatemi}, E. 1992,
  \href{http://dx.doi.org/10.1016/0167-2789(92)90242-F}{\JournalTitle{Physica D
  Nonlinear Phenomena}, 60, 259}

\bibitem[{{Schwarzschild}(1975)}]{S75}
{Schwarzschild}, M. 1975,
  \href{http://dx.doi.org/10.1086/153313}{\JournalTitle{\apj}, 195, 137}

\bibitem[{{Selina} {et~al.}(2018){Selina}, {Murphy}, {McKinnon}, {Beasley},
  {Butler}, {Carilli}, {Clark}, {Durand}, {Erickson}, {Grammer}, {Hiriart},
  {Jackson}, {Kent}, {Mason}, {Morgan}, {Ojeda}, {Rosero}, {Shillue},
  {Sturgis}, \& {Urbain}}]{Selina2018}
{Selina}, R.~J., {Murphy}, E.~J., {McKinnon}, M., {et~al.} 2018, in
  Astronomical Society of the Pacific Conference Series, Vol. 517, Science with
  a Next Generation Very Large Array, ed. E.~{Murphy}, 15

\bibitem[{{Vlemmings} {et~al.}(2017){Vlemmings}, {Khouri}, {O'Gorman}, {De
  Beck}, {Humphreys}, {Lankhaar}, {Maercker}, {Olofsson}, {Ramstedt}, {Tafoya},
  \& {Takigawa}}]{Vlemmings2017}
{Vlemmings}, W., {Khouri}, T., {O'Gorman}, E., {et~al.} 2017,
  \href{http://dx.doi.org/10.1038/s41550-017-0288-9}{\JournalTitle{Nature
  Astronomy}, 1, 848}

\bibitem[{{Vlemmings} {et~al.}(2019){Vlemmings}, {Khouri}, \&
  {Olofsson}}]{Vlemmings2019}
{Vlemmings}, W.~H.~T., {Khouri}, T., \& {Olofsson}, H. 2019,
  \href{http://dx.doi.org/10.1051/0004-6361/201935329}{\JournalTitle{\aap},
  626, A81}

\end{thebibliography}
\bibliographystyle{yahapj}

\end{document}